\documentstyle[psfig]{mn}

\title[Optical spectroscopy of faint GPS sources]{Optical spectroscopy of
faint gigahertz peaked spectrum sources}

\author[I. Snellen et al.]{I.A.G. Snellen$^{1,2}$, R.T. Schilizzi$^{1,3}$,
M.N. Bremer$^{1,6,7}$, G.K. Miley$^{1}$, A.G. de Bruyn$^{4,5}$, \cr
H.J.A. R\"ottgering$^{1}$\\ 
$^{1}$Leiden Observatory, P.O. Box 9513, 2300 RA, Leiden, The Netherlands \\
$^{2}$Institute of Astronomy, Madingley Road, Cambridge CB3 0HA, United 
Kingdom\\
$^{3}$Joint Institute for VLBI in Europe, Postbus 2, 7990 AA, Dwingeloo, 
The Netherlands\\
$^{4}$Netherlands Foundation for Research in Astronomy, Postbus 2, 7990 AA, 
Dwingeloo, The Netherlands\\
$^{5}$Kapteyn Institute, Postbus 800, 9700 AV, Groningen, The Netherlands\\
$^{6}$Institut d'Astrophysique de Paris, 98bis Boulevard Arago, 75014 
         Paris, France\\
$^{7}$ Department of Physics, Bristol University, H H Wills Physics Laboratory,  Tyndall Avenue, Bristol, BS8 1TL, United Kingdom}

\date{}
\begin{document}
\maketitle

\begin{abstract}
We present spectroscopic observations of a sample of faint Gigahertz Peaked
Spectrum (GPS) radio sources drawn from the Westerbork Northern Sky Survey
(WENSS). Redshifts have been determined for 19 (40\%) of the objects.
The optical spectra of the GPS sources identified with low redshift
galaxies show deep stellar absorption features. 
This confirms previous suggestions that their optical light is not 
significantly contaminated by AGN-related emission,
but is dominated by a population of old ($>$9 Gyr) and metal-rich 
($>$0.2 $[$Fe/H$]$) stars, justifying the use of these (probably) young 
radio sources as probes of galaxy evolution.
The optical spectra of GPS sources identified with quasars are 
indistinguishable from those of flat spectrum quasars, and clearly different 
from the spectra of Compact Steep Spectrum (CSS) quasars. 
The redshift distribution of the GPS quasars in our radio-faint sample is 
comparable to that of the bright samples presented in the literature, peaking 
at $z\sim 2-3$. It is unlikely that a significant population of low redshift 
GPS quasars is missed due to selection effects in our sample. 
We therefore claim that 
there is a genuine difference between the redshift distributions of 
GPS galaxies and quasars, which, because it is present in both the 
radio-faint and bright samples, can not be due to a redshift-luminosity
degeneracy. It is therefore unlikely that the GPS quasars and galaxies are 
unified by orientation, unless the quasar opening angle is a strong function 
of redshift. We suggest that the GPS quasars and galaxies are unrelated 
populations and just happen to have identical observed radio-spectral 
properties, and hypothesise that GPS quasars are a sub-class of flat spectrum 
quasars.
\end{abstract}

\section{Introduction}

Gigahertz Peaked Spectrum (GPS) sources are a class of compact radio
source, characterised by a convex radio spectrum peaking at a
frequency of $\sim$ 1 GHz. Together with the class of Compact Steep
Spectrum (CSS) sources, which have spectra peaking at lower
frequencies, they form a significant fraction ($\sim$ 30\%) of the 
high frequency radio source population (see O'Dea 1998, for a recent
review on GPS and CSS sources).  Their optical counterparts are a
mixture of quasars and galaxies, with the quasars preferentially found
at significantly higher redshifts than the galaxies 
($z\sim 2-3$, O'Dea et al 1991, Stanghellini et al. 1998).  
The VLBI-morphologies (Stanghellini et al. 1997) and
the distributions of radio spectral peak frequencies (de Vries et al.,
1997; Snellen et al., 1998b) also seem to be different for GPS
galaxies and quasars. These differences cause severe problems for 
orientation-based unification of 
GPS galaxies and quasars.

Their small sizes ($\sim$ 100 pc) have made GPS sources the prime 
candidate to represent the early stages of radio source evolution
(Fanti et al. 1995, Readhead et al. 1996, O'Dea and Baum 1997).
The alternative hypothesis, which assumes these sources to be small due
to confinement by a particularly dense interstellar medium (O'Dea et al. 
1991), is less likely as recent observations show that the 
surrounding media of GPS sources are not significantly different from 
large scale sources (eg. Fanti et al. 1995). In addition, Owsianik and Conway 
(1998) have determined that the dynamical age of the radio source 0710+319, 
a prototype GPS galaxy,
to be  $\sim$ $10^3$ years by measuring the propagation velocity of its hot 
spots. It is therefore likely that at least some of the GPS galaxies
represent the young counterparts of ``old'' extended radio sources. 
These are therefore the objects of choice to study the initial evolution
of radio sources. Investigating the optical hosts and environments 
of GPS sources provides important information on the circumstances 
under which a radio source is formed.

Here we present results on a sample of faint GPS sources
(Snellen 1997) 
selected from the Westerbork Northern Sky Survey (WENSS; 
Rengelink et al. 1997).
The combination of this new faint sample and existing brighter samples 
(Fanti et al 1990; O'Dea et al 1991; Stanghellini et al., 1998; 
de Vries et al., 1997) allow for the first time the disentanglement of 
redshift and radio luminosity effects.

Previous papers have described the selection 
(Snellen et al. 1998a), and optical and near-infrared imaging of the 
radio-faint GPS sample (Snellen et al. 1998b). This paper describes the 
spectroscopic observations, leading to a discussion of the emission and 
absorption line properties and the redshift distributions of the GPS galaxies 
and quasars.

\section{The Sample}

The selection of the sample has been described in  detail in Snellen et al. 
(1998a).
Candidate GPS sources selected from the Westerbork Northern Sky survey, are
those with an inverted spectrum between 325 MHz and higher frequencies.
The sources are located in two regions of the survey; one at $15^h < \alpha < 
20^h$
and $58^\circ< \delta < 75^\circ$, which is called the {\it mini-survey} region
(Rengelink et al. 1997), and the other at $4^h00^m < \alpha < 8^h30^m$ and
$58^\circ< \delta < 75^\circ$. Additional observations at 1.4, 5, 8.4 and 15
GHz were carried out with the WSRT and the VLA, yielding a sample of 47
genuine GPS sources with peak frequencies ranging from 500 MHz to more than 15
GHz, and peak flux densities ranging from $\sim30$ to $\sim900$ mJy.
This sample has been imaged in the optical and near-infrared, resulting in
an identification fraction of $\sim$ 87 \% (Snellen et al. 1998b).

The subsample for which we
obtained optical spectra was selected in the following way. During the first
observing session with the William Herschel Telescope (WHT), 5 of the 15 GPS 
sources in the mini-survey region
identified with $R>22$ GPS objects were observed, but no redshifts could be
determined due to the low signal-to-noise ratio of these observations. 
Subsequently, we concentrated on the brighter sources in
the sample. In the mini-survey region the 11 sources identified with an 
$R<22.0$ object were observed. 
In the region between $4^h00^m < R.A. < 8^h30^m$,
the eight sources identified with an $R<20.0$ object were observed, except for
B0441+5757 which could not be observed due to scheduling constraints.

\section{Observations and Reduction}

\begin{table}
\caption{\label{obs} Log of the Observations. Column 1 gives the sourcename,
column 2 the $R$ band magnitude from Snellen et al. (1998b), column 3 
the telescope used, column 4 the date of observations, 
and column 5 the exposure time.}
\centerline{
\begin{tabular}{cccrr}\hline
      &     &         &    &\\
Object&$m_R$&Tel.&Date        &Exp.\\
      &(mag)&            &            &  (sec)  \\ \hline
B0531+6121&19.0&WHT&20 Dec. 1995&1800\\
B0537+6444&19.5&WHT& 8 Jan. 1997&1800\\
B0544+5847&19.4&WHT& 8 Jan. 1997&1800\\
B0601+5753&19.1&WHT&20 Dec. 1995&1800\\
B0755+6354&19.1&WHT&20 Dec. 1995&1800\\
B0758+5929&19.3&WHT& 8 Jan. 1997&1800\\
B0826+7045&19.7&WHT& 8 Jan. 1997&1800\\
B0830+5813&15.9&INT&11 Nov. 1996& 600\\
B1525+6801&23.1&WHT&25 Jul. 1995&3000\\
B1538+5920&20.9&WHT&20 Jun. 1997&1800\\
B1550+5815&16.7&INT&30 Jul. 1995&1200\\
B1551+6822&23.8&WHT&26 Jul. 1995&3000\\
B1557+6220&22.5&WHT&24 Jul. 1995&3000\\
B1622+6630&17.2&WHT&12 Aug. 1996&3000\\
B1642+6701&17.0&INT&30 Jul. 1995&1200\\
B1647+6225&22.9&WHT&12 Aug. 1996&3000\\
B1746+6921&19.2&INT&30 Jul. 1995&1800\\
B1819+6707&17.7&INT&30 Jul. 1995&2400\\
B1841+6715&20.5&WHT&31 Aug. 1994&3600\\
B1942+7214&23.0&WHT&26 Jul. 1995&3000\\
B1945+6024&20.4&INT&30 Jul. 1995&3000\\
B1946+7048&16.3&INT&30 Jul. 1995& 900\\
B1954+6146&22.2&WHT&25 Jul. 1995&3000\\
B1958+6158&19.1&WHT&19 Jun. 1997&1800\\ \hline
\end{tabular}}
\end{table}

\begin{table*}
\setlength{\tabcolsep}{2.0mm}
\caption{\label{lines}Emission line properties of GPS sources.
column 1 gives the source name, column 2 the identification (Q/G), column 3 
the observed wavelengths of the 
emission lines, column 4 the redshifts, column 5 the line flux, column
6 the FWHM, and column 7 the rest-frame equivalent width.}
\renewcommand{\arraystretch}{0.8}
\begin{tabular}{|c|ccclcrr|}\hline
Object&ID&Line&Peak &Redshift&Flux               &FWHM& Eq. Width\\
      &   &    &$\AA$&              &$10^{-16}\frac{erg}{cm^2 \ s}$ &$km/s$&(rest)  $\AA$\\ \hline
B0531+6121&G&    &               &$0.414\pm0.0010$&            &           &\\
         &&$[$OII$]$ &$5266.9\pm0.9 $&$0.413\pm0.0002$&$24.2\pm2.4$&$<382$&$50\pm6     $\\
         &&H$\beta$  &$6875.5\pm0.6 $&$0.414\pm0.0001$&$ 12.4\pm1.4$&$<349$&$15\pm2$\\
         &&$[$OIII$]$&$7016.1\pm0.6 $&$0.415\pm0.0001$&$48.5\pm8.0$&$<285$&$44\pm5$\\
         &&$[$OIII$]$&$7084.1\pm0.5 $&$0.415\pm0.0001$&$128.3\pm14.4$&$604\pm280$&$160\pm16$\\\hline
B0537+6444&Q&    &               &$2.417\pm0.0010$&            &           &\\
         &&Ly$\alpha$ &$4154.6\pm 1.2$&$2.417\pm0.0010$&$44.2\pm6.0$&$2550\pm 250$&$   159 \pm  16$\\  
       &&SiIV/O&$4782.6\pm31.1$&$2.416\pm0.0222$&$ 3.0\pm1.0$&$3580\pm700$&$  10\pm2$\\    
         &&CIV &$5292.8\pm 1.8$&$2.417\pm0.0012$&$17.3\pm2.7$&$2800\pm 300$&$   50\pm   5$\\    
         &&CIII$]$&$6509.7\pm 7.7$&$2.410\pm0.0040$&$ 8.6\pm1.2$&$2150\pm 200$&$   23 \pm   2$\\\hline
         
B0544+5847&    &               &$2.860\pm0.0070$&            &           &\\
         &Q&Ly$\alpha$ &$4780.5\pm151.9$&$2.931\pm0.1250$&$15.0\pm2.3$&$13500\pm1500$&$17\pm2$\\ 
         &&CIV &$5977.5\pm 10.8$&$2.859\pm0.0070$&$ 4.5\pm0.9$&$ 5350\pm 1000$&$7\pm1$\\
         &&CIII$]$&$7350.0\pm 26.9$&$2.850\pm0.0141$&$ 4.0\pm0.8$&$ 3300\pm600$&$7\pm1$\\\hline       
B0601+5753&Q&    &               &$1.840\pm0.0020$&                   &      &\\
         &&CIV &$4402.5\pm3.8$ &$1.842\pm0.0025$&$30.0\pm5.8$&$3650\pm400$&$42\pm4$\\
         &&HeII&$4652.8\pm4.7$ &$1.837\pm0.0029$&$ 3.0\pm0.7$&$1450\pm300$&$ 6\pm1$\\
       &&CIII$]$  &$5425.0\pm10.5$&$1.842\pm0.0055$&$4.5\pm1.1$&$2650\pm500$&$10\pm2$\\   
         &&MgII&$7992.6\pm59.8$&$1.856\pm0.0212$&$ 7.4\pm2.0$&$4850\pm900$&$26\pm3$\\ \hline
B0755+6354&Q&    &               &$3.005\pm0.0010$&            &             &        \\
        &&Ly$\alpha$  &$4872.7\pm1.3 $&$3.007\pm0.0011$&$148.2\pm14$&$1750\pm200$&$110\pm9$\\  
        &&CIV  &$6194.4\pm2.6 $&$2.999\pm0.0017$&$53.3\pm5.3$&$3600\pm350$&$57\pm6$\\\hline   

B0758+5929&Q&    &               &$1.977\pm0.0190$&            &            &          \\
        &&CIV  &$4612.2 \pm 29.8$&$1.978\pm0.0193$&$8.9\pm0.9$&$6400\pm 650$&$    25\pm3$\\
        &&CIII$]$ &$5652.7 \pm179.5$&$1.961\pm0.0946$&$5.1\pm0.5$&$9350\pm 1000$&$  14\pm2$\\\hline

B0826+7045&Q&    &               &$2.003\pm0.0010$&           &              &        \\
      &&SiIV/O &$4204.5\pm87.4$&$2.003\pm0.0624$&$ 8.5\pm0.8$&$ 7950\pm 1800$&$   18 \pm 5$\\
        &&CIV  &$4659.7\pm 6.9$&$2.008\pm0.0044$&$39.9\pm4.0$&$ 6150\pm  600$&$   42 \pm 5  $\\
        &&HeII &$4919.3\pm 2.5$&$2.000\pm0.0015$&$ 3.6\pm0.4$&$   -^a $       &$    4 \pm 1 $ \\
        &&CIII$]$ &$5720.6\pm 2.1$&$1.997\pm0.0011$&$32.4\pm3.2$&$ 4000\pm  400$&$   38 \pm   4 $\\\hline
B0830+5813&G&    &               &$0.093\pm0.0007^b$&            &               &                \\
        &&NII   &$7188.3\pm 4.3$&$0.093\pm0.0007$&$16.2\pm1.6$&$ 1796\pm  434$&$   15 \pm   2 $\\\hline
B1538+5920&Q&    &               &$3.878\pm0.0010$&            &            &          \\
         &&Ly$\alpha$ &$5929.2\pm1.2 $&$3.876\pm0.0009$&$11.2\pm0.8$&$1900\pm200$&$37\pm4 $\\
         && NV &$6054.6\pm1.6 $&$3.883\pm0.0013$&$3.4\pm0.3$&$2700\pm270      $&$13\pm1   $\\
         &&CIV &$7544.2\pm3.9 $&$3.870\pm0.0025$&$9.7\pm0.4$&$2500\pm250$&$35\pm10     $\\\hline
B1550+5815&Q&    &               &$1.324\pm0.0020$&           &            &           \\
         &&CIII$]$&$4427.7\pm5.3 $&$1.319\pm0.0028$&$395.1\pm40$&$4550\pm500$&$20\pm2$\\  
         &&MgII&$6509.0\pm5.2 $&$1.325\pm0.0019$&$ 320.7\pm32$&$3300\pm350$&$31\pm3$\\ \hline
B1622+6630&G&    &               &$0.201\pm0.0003$&              &            &          \\
         &&$[$OII$]$ &$4490.9\pm2.0 $&$0.205\pm0.0015$&$ 7.3\pm1.1$&$<400$&$18\pm3$\\
         &&H$\beta$  &$5835.4\pm0.8 $&$0.200\pm0.0002$&$ 4.6\pm0.5$&$<267$&$6\pm1$\\
         &&$[$OIII$]$&$5958.8\pm0.8 $&$0.202\pm0.0002$&$21.1\pm2.2$&$<300$&$10\pm2$\\ 
         &&$[$OIII$]$&$6016.6\pm0.8 $&$0.202\pm0.0002$&$51.7\pm5.2$&$<330$&$25\pm3$\\  
         &&H$\alpha$  &$7895.2\pm2.9 $&$0.203\pm0.0004$&$312.0\pm31.5$&$3540\pm347$&$114\pm12$\\ \hline
B1642+6701&Q&    &               &$1.895\pm0.0010$&            &           &\\
       &&SiIV/O&$4055.4\pm2.4 $&$1.897\pm0.0017$&$158.0\pm16$&$5550\pm1100$&$7\pm1$\\
         && CIV&$4484.0\pm1.6 $&$1.895\pm0.0010$&$469.0\pm64$&$6000\pm600$&$27\pm3$\\
         &&CIII$]$&$5505.9\pm20.1$&$1.884\pm0.0105$&$263.0\pm26$&$5300\pm500$&$21\pm2$\\ \hline  
B1647+6225&Q&    &               &$2.167\pm0.0010$&              &             &\\
         &&Ly$\alpha$ &$3869.7\pm2.0 $&$2.182\pm0.0019$&$9.8\pm1.0$&$2850\pm 300$&$181\pm18$\\
         &&NV  &$3942.4\pm5.7 $&$2.179\pm0.0046$&$3.5\pm0.4$&$3400\pm350$&$69\pm7$\\
         &&CIV &$4926.3\pm1.9 $&$2.180\pm0.0012$&$4.3\pm0.4$&$2700\pm300$&$126\pm25$\\
         &&CIII$]$&$6063.9\pm7.6 $&$2.176\pm0.0044$&$1.1\pm0.2$&$2200\pm 250$&$32\pm3$\\ \hline 
\end{tabular}

$^a$ No reliable determination of FWHM posibble due to associated absorption.\\
$^b$ several stellar absorption lines are present consistent with the emission
line redshift.\\

\end{table*}
\addtocounter{table}{-1}
\begin{table*}
\setlength{\tabcolsep}{2.0mm}
\renewcommand{\arraystretch}{0.8}
\begin{tabular}{|c|cccccrr|}\hline
Object&ID&Line&Peak &Redshift&Flux               &FWHM& Eq. Width\\
      &   &    &$\AA$&              &$10^{-16}\frac{erg}{cm^2 \ s}$ &$km/s$&(rest)  $\AA$\\ \hline
B1746+6921&Q&    &               &$1.886\pm0.0040$&           &             &\\
       &&   CIV&$4468.9\pm 8.4$&$1.885\pm0.0054$&$20.8\pm2.1$&$2500\pm500$&$9\pm1$\\
         &&CIII$]$&$5514.4\pm14.4$&$1.889\pm0.0075$&$32.0\pm3.2$&$5650\pm550$&$17\pm2$\\\hline
B1819+6707&G&    &               &$0.220\pm0.0003$&            &             &\\
        &&$[$OII$]$ &$4546.6\pm1.2$ &$0.220\pm0.0003$&$5.7\pm0.7$&$<442$&$22\pm3$\\ 
        &&$[$OIII$]$ &$6047.1\pm4.8$ &$0.219\pm0.0010$&$2.2\pm0.4$&$<615$&$3\pm1$\\  
        &&$[$OIII$]$ &$6110.2\pm1.8$ &$0.220\pm0.0004$&$4.6\pm0.6$&$<372$&$6\pm2$\\  \hline
B1841+6715&G&    &               &$0.470\pm0.0100$&           &      &\\ 
        &&$[$OII$]$  &$5459.1\pm1.1 $&$0.465\pm0.0003$&$1.9\pm0.3$&$<366$&$13\pm2$\\
        &&H$\beta$   &$7220.8\pm2.0 $&$0.485\pm0.0004$&$0.4\pm0.1$&$<300$&$2\pm1$\\ 
        &&$[$OIII$]$ &$7433.3\pm1.9 $&$0.485\pm0.0004$&$3.9\pm0.5$&$317\pm305$&$23\pm2$\\ \hline
B1945+6024&Q&    &               &$2.700\pm0.0080$&           &      &\\
        &&Ly$\alpha$  &$4519.0\pm13.0$&$2.716\pm0.0106$&$3.3\pm0.3$&$7150\pm1400$&$12\pm2$\\
        &&CIV  &$5724.6\pm 8.1$&$2.696\pm0.0052$&$1.9\pm0.2$&$3900\pm800$&$7\pm1$\\\hline
B1946+7048&G&    &               &$0.101\pm0.0005$&           &           &\\
        &&$[$OII$]$  &$4107.3\pm1.3$ &$0.102\pm0.0003$&$ 7.6\pm2.3$&$<503     $&$24\pm3$\\
        && H$\beta$  &$5353.7\pm1.3$ &$0.101\pm0.0003$&$ 5.4\pm0.8$&$<388     $&$4\pm0$\\
        &&$[$OIII$]$ &$5460.1\pm2.6$ &$0.101\pm0.0005$&$ 4.0\pm1.5$&$<400     $&$3\pm1$\\
        &&$[$OIII$]$ &$5514.4\pm1.2$ &$0.101\pm0.0002$&$ 9.2\pm1.5$&$<366     $&$7\pm1$\\
        &&$[$OI$]$   &$6935.4\pm2.0$ &$0.101\pm0.0003$&$ 8.0\pm1.1$&$<306     $&$5\pm1$\\\hline
B1958+6158&Q&    &               &$1.824\pm0.0010$&            &           &\\
        &&Ly$\alpha$  &$4628.5\pm1.3 $&$1.823\pm0.0008$&$106.0\pm15$&$3350\pm350$&$ 81\pm8$\\
       &&SiIV/O&$3952.8\pm6.9 $&$1.823\pm0.0049$&$23.0\pm3.0$&$5500\pm550$&$ 18\pm1$\\
        &&CIV  &$4373.8\pm1.5 $&$1.824\pm0.0009$&$57.0\pm7.0$&$2650\pm300$&$  37\pm4$\\
        &&HeII &$4629.5\pm1.6 $&$1.823\pm0.0010$&$ 4.0\pm0.6$&$2200\pm500$&$   3\pm1$\\
        &&CIII$]$ &$5382.4\pm8.3 $&$1.819\pm0.0043$&$18.0\pm2.4$&$5650\pm550$&$  17\pm2$\\
        &&MgII &$7907.8\pm2.6 $&$1.825\pm0.0009$&$9.0\pm1.4$&$1400\pm300$&$  11\pm2$\\\hline
\end{tabular}
\caption{Continued...}
\end{table*}

The observations were carried out using the 2.5m Isaac Newton Telescope (INT)
and the 4.2m William Herschel telescope (WHT). Some of
these observations were done as part of a programme carried out in the 
international CCI observing period on the La Palma telescopes in 1995. 
Table \ref{obs} gives the log of the
observations. 
The ISIS long slit spectrograph was used for the WHT observations, with
a $1024\times 1024$ Tektronix CCDs in both the red and blue arms. A spectral 
resolution of $\sim 15 $\AA \ was
obtained between 3500 and 9000 \AA , using a slit width of 2$''$
 and R158 gratings. The Intermediate Dispersion Spectrograph (IDS) was
used for the INT observations, with a Tektronix CCD. A spectral resolution 
of $\sim 10 $\AA \ was obtained between 4000 and 7500 \AA \ using a slit width of two arcsec and a grating with 300 grooves/mm. Usually the slit was
oriented near the paralactic angle.

The reduction of the spectra was carried out using the `Long Slit' package of
the NOAO's IRAF reduction software. 
A bias frame was constructed by averaging
`zero second' exposures taken at the beginning of each night. This was
subtracted from every non-bias frame. The pixel-to-pixel variations were
calibrated using flat fields obtained from an internal quartz lamp. Wavelength
calibration was carried out by measuring the positions on the frames of known
lines from either an Cu-Ne or a Cu-Ar calibration lamp. The sky contribution
was removed by subtracting a sky spectrum obtained by fitting a polynomial to
the intensities measured along the spatial direction outside the vicinity of
targets. One dimensional spectra were extracted by averaging in the spatial
direction over an aperture as large as the spatial extent of the brightest
emission line.

\section{Results}

Redshifts were determined for all 11 sources in the mini-survey region
identified with an $R<22.0$ object, and for 8 sources in the region between
$4^h00^m < R.A. < 8^h30^m$, which are all sources in this region identified
with an $R<20.0$ object, except B0441+5757. Five of the 15 GPS sources in the
mini-survey region identified with $R>22$ GPS objects were observed, but no
redshifts could be determined: For four objects no lines were seen at a
level of $>2 \times 10^{16}$ and the 
signal-to-noise ratio was too poor to detect absorption against the faint 
continuum. In addition, there were no spectral features observed in B1954+6154
to determine the redshift (see below).
The spectra of the 19 sources with determined redshifts and of B1954+6154 are
shown in figure \ref{optspec}. Their emission line properties are given in 
table \ref{lines}.
The final redshifts of the objects and their uncertainties were determined,
 taking into account the 
redshift differences and uncertainties of the individual lines.
A distinction between galaxies and quasars has been made on grounds of 
optical morphology (stellar-extended) and emission line widths. 
A distinction between galaxies and quasars has been made on grounds of 
optical morphology (stellar-extended) and emission line widths. 
The optical host is called a quasar when there is no resolved galaxy visible
in the optical CCD-image (Snellen et al. 1998b) and the spectrum shows emission
lines with FWHM $>$ 2500 km sec$^{-1}$.

One of the objects, B1954+6146, has a very peculiar spectrum. Below 6000 \AA, 
the spectrum is faint and flat, while redward of 6000\AA \ a steep rise in 
luminosity occurs. It was 
therefore not clear whether we should classify this object as a galaxy or
a quasar. No lines could be identified. 
 The $R$ band image of this object 
(Snellen et al. 1998b) shows a
point source surrounded by faint extended emission. It is likely that
the optical morphology and spectrum are produced by a non-thermal quasar 
nucleus visible in the red, but obscured in the blue. 
This would imply a reddening of about 9-10 magnitudes in the $R$ band, 
assuming a flat spectral energy distribution for the quasar nucleus and a 
standard $1/\lambda$ extinction curve.

\begin{figure*}
\hbox{
\psfig{figure=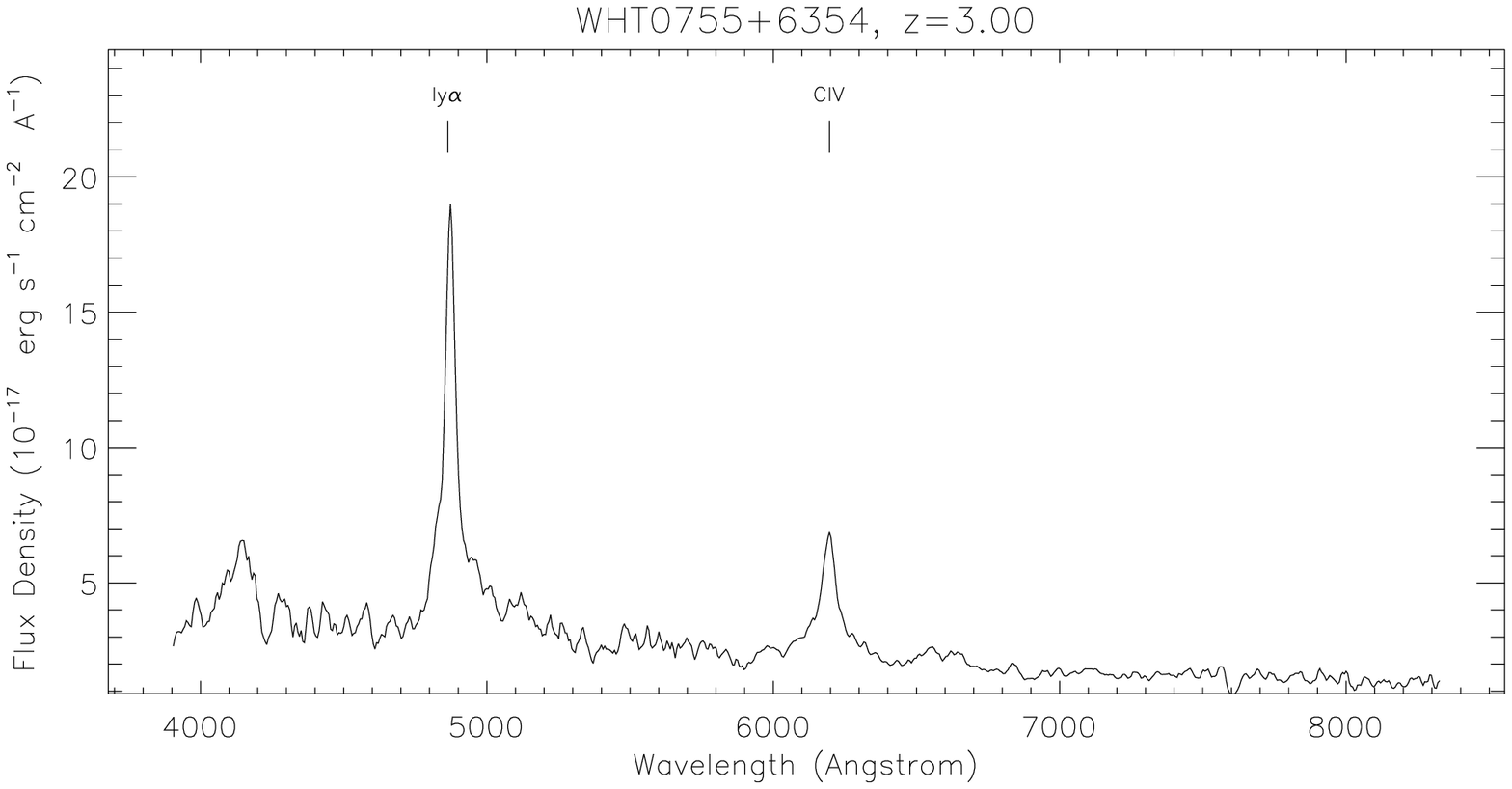,width=3.75cm,height=10.75cm,angle=90}
\psfig{figure=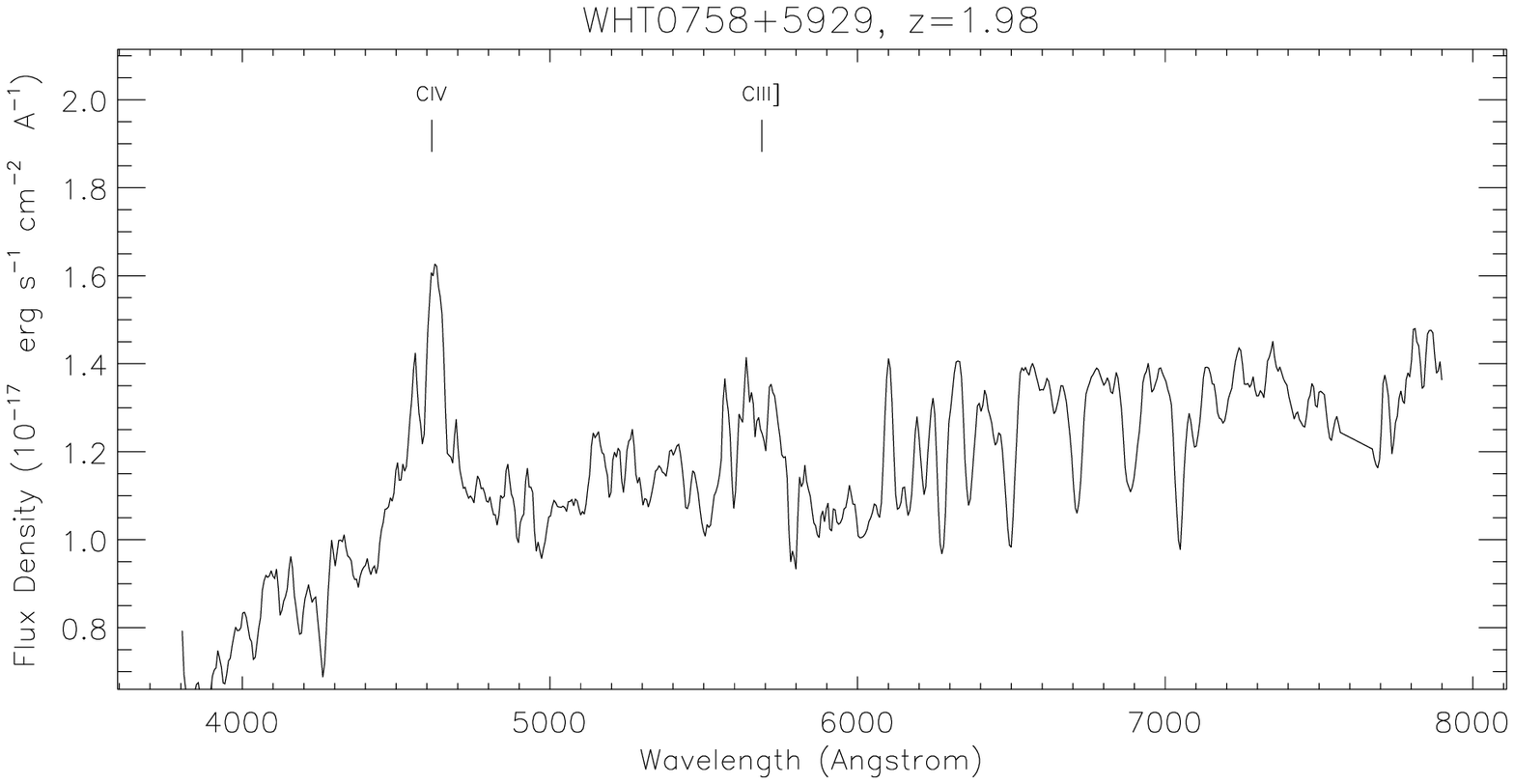,width=3.75cm,height=10.75cm,angle=90}
\psfig{figure=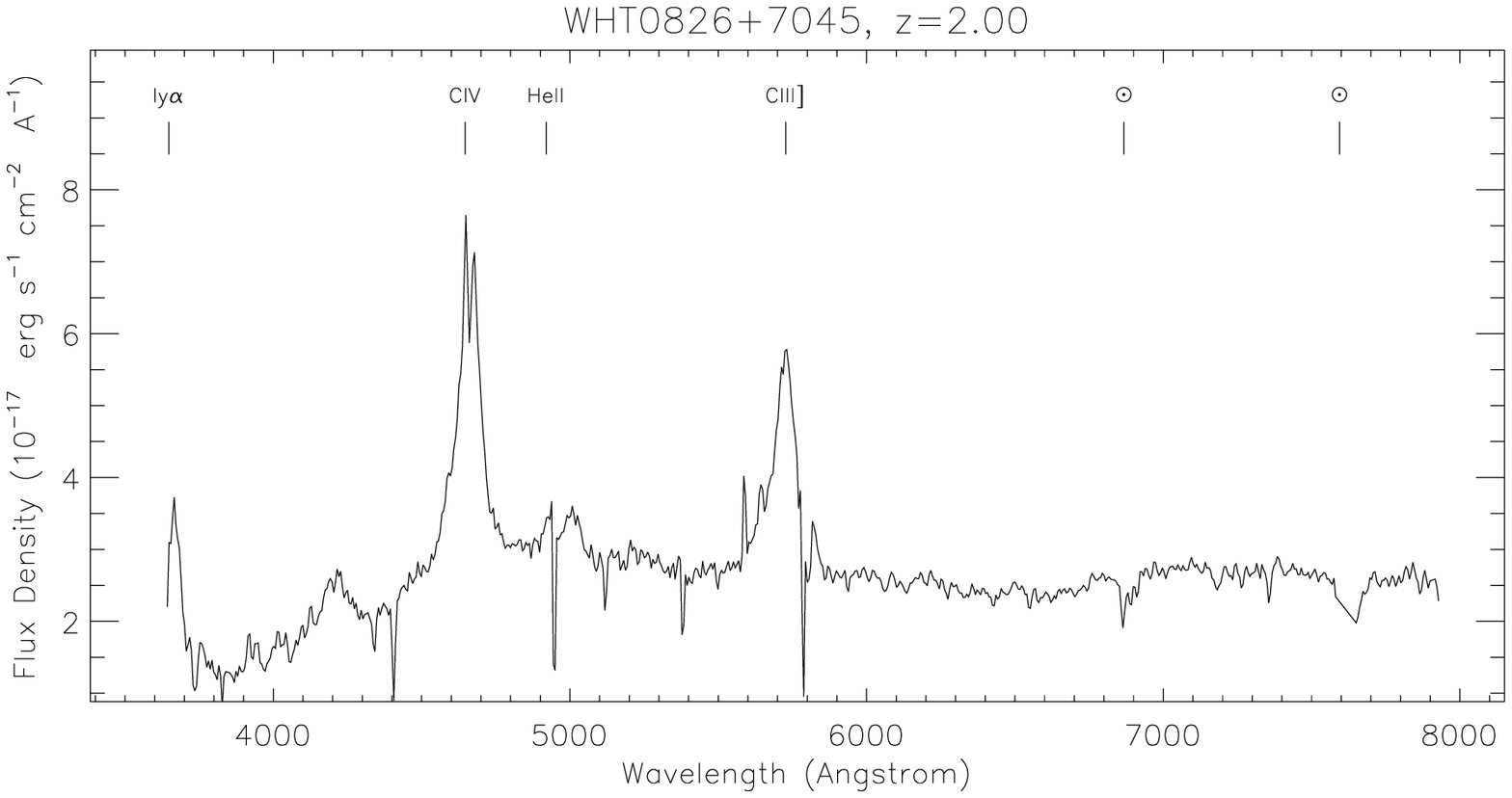,width=3.75cm,height=10.75cm,angle=90}
\psfig{figure=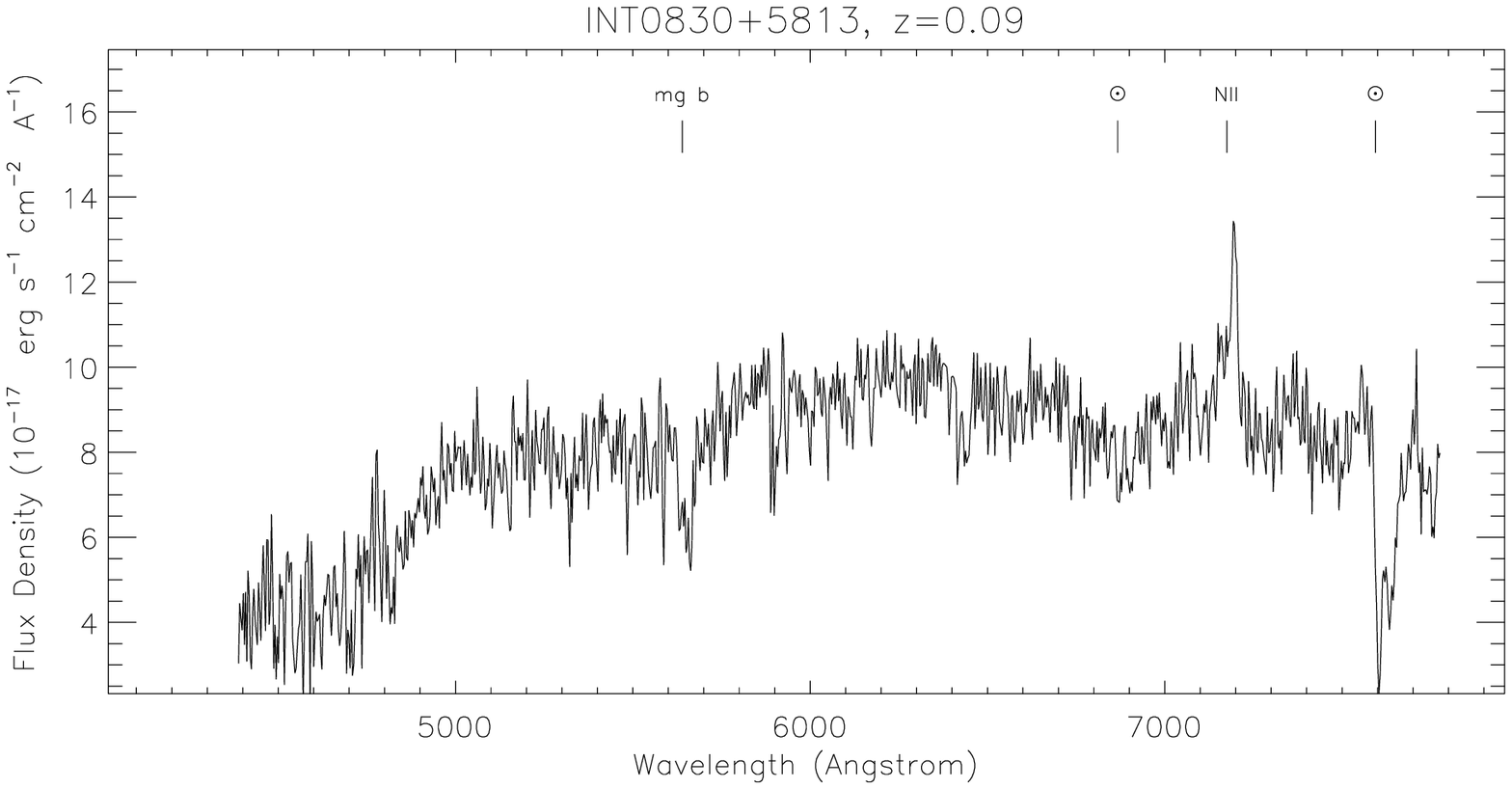,width=3.75cm,height=10.75cm,angle=90}}
\hbox{
\psfig{figure=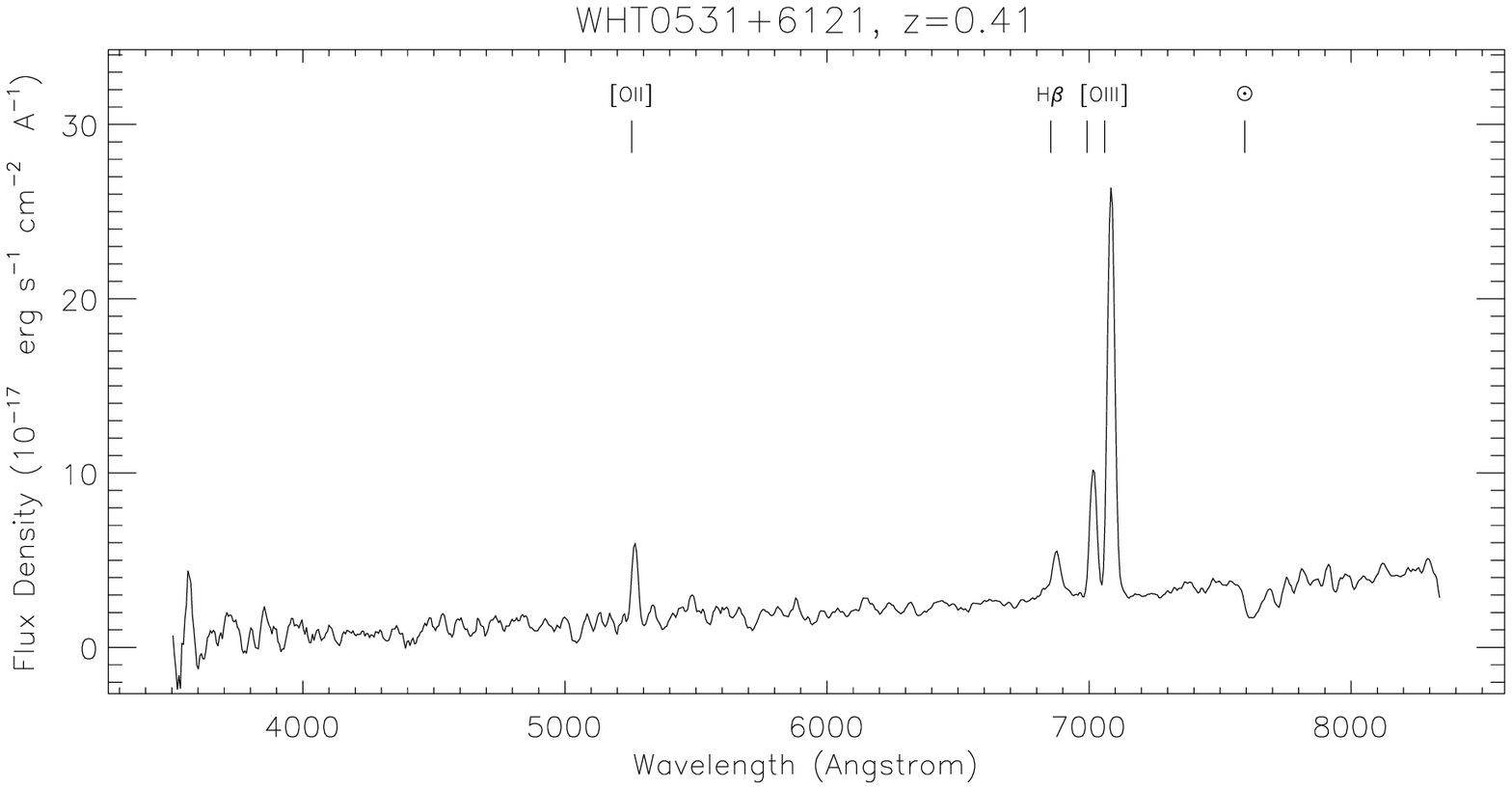,width=3.75cm,height=10.75cm,angle=90}
\psfig{figure=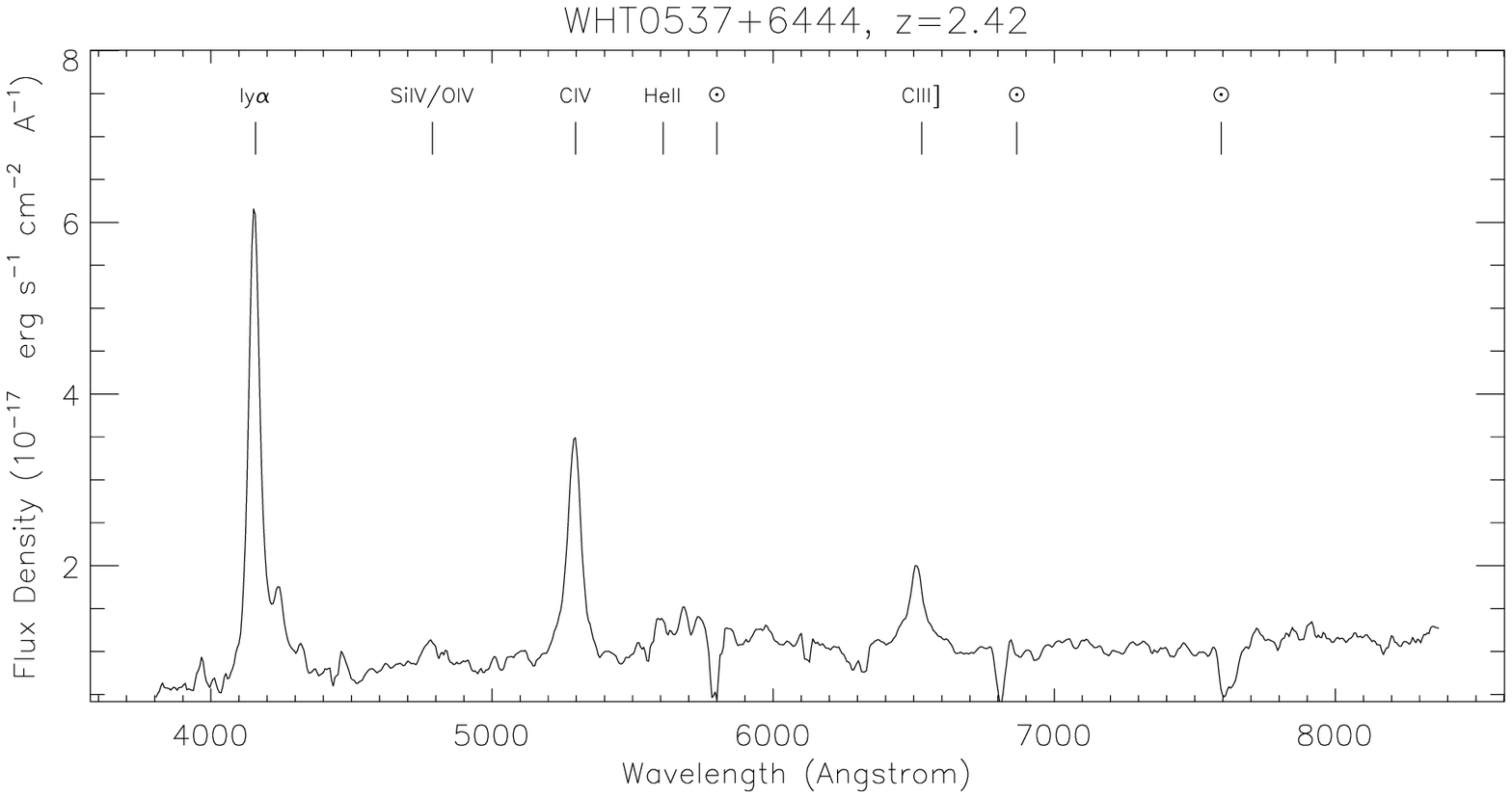,width=3.75cm,height=10.75cm,angle=90}
\psfig{figure=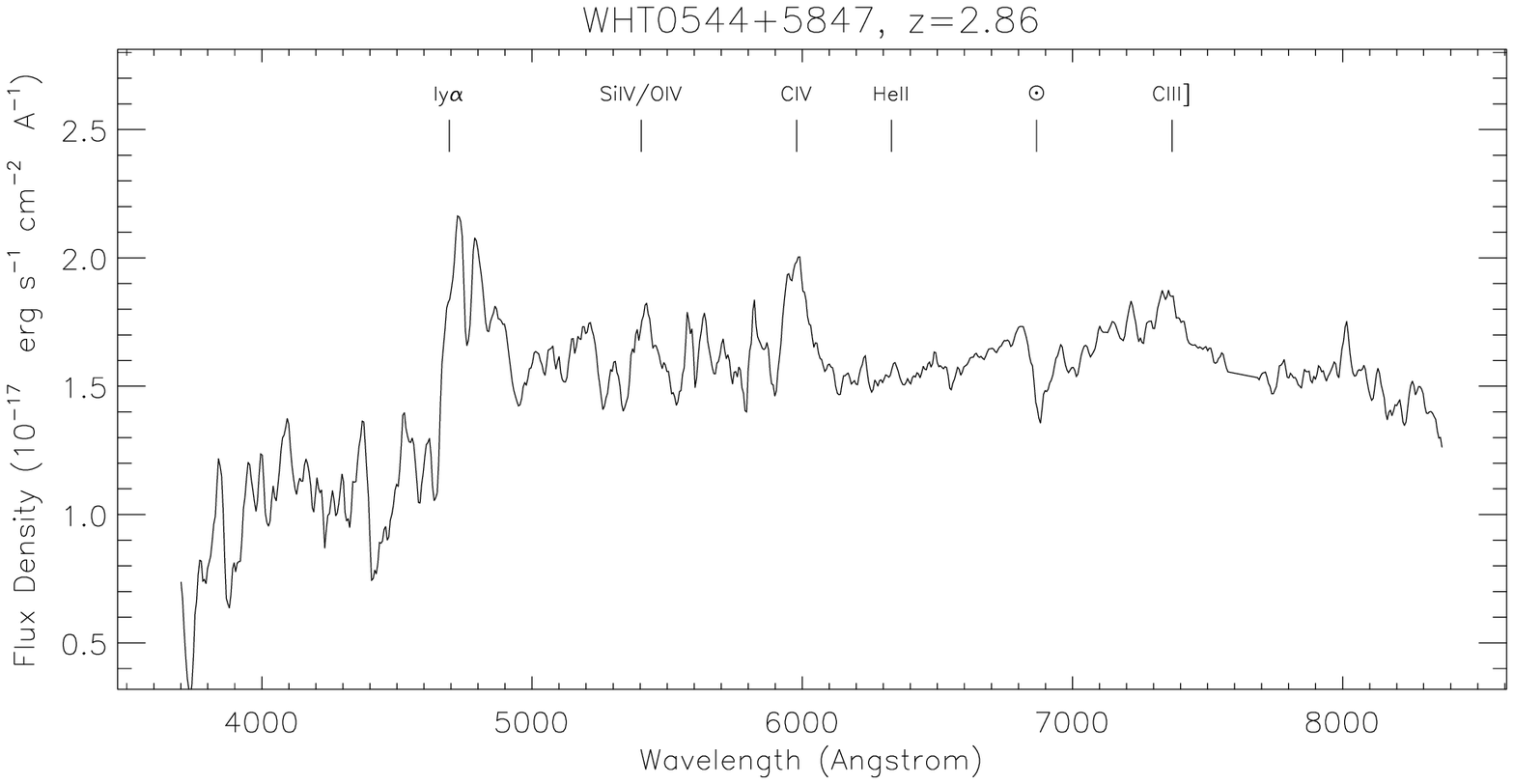,width=3.75cm,height=10.75cm,angle=90}
\psfig{figure=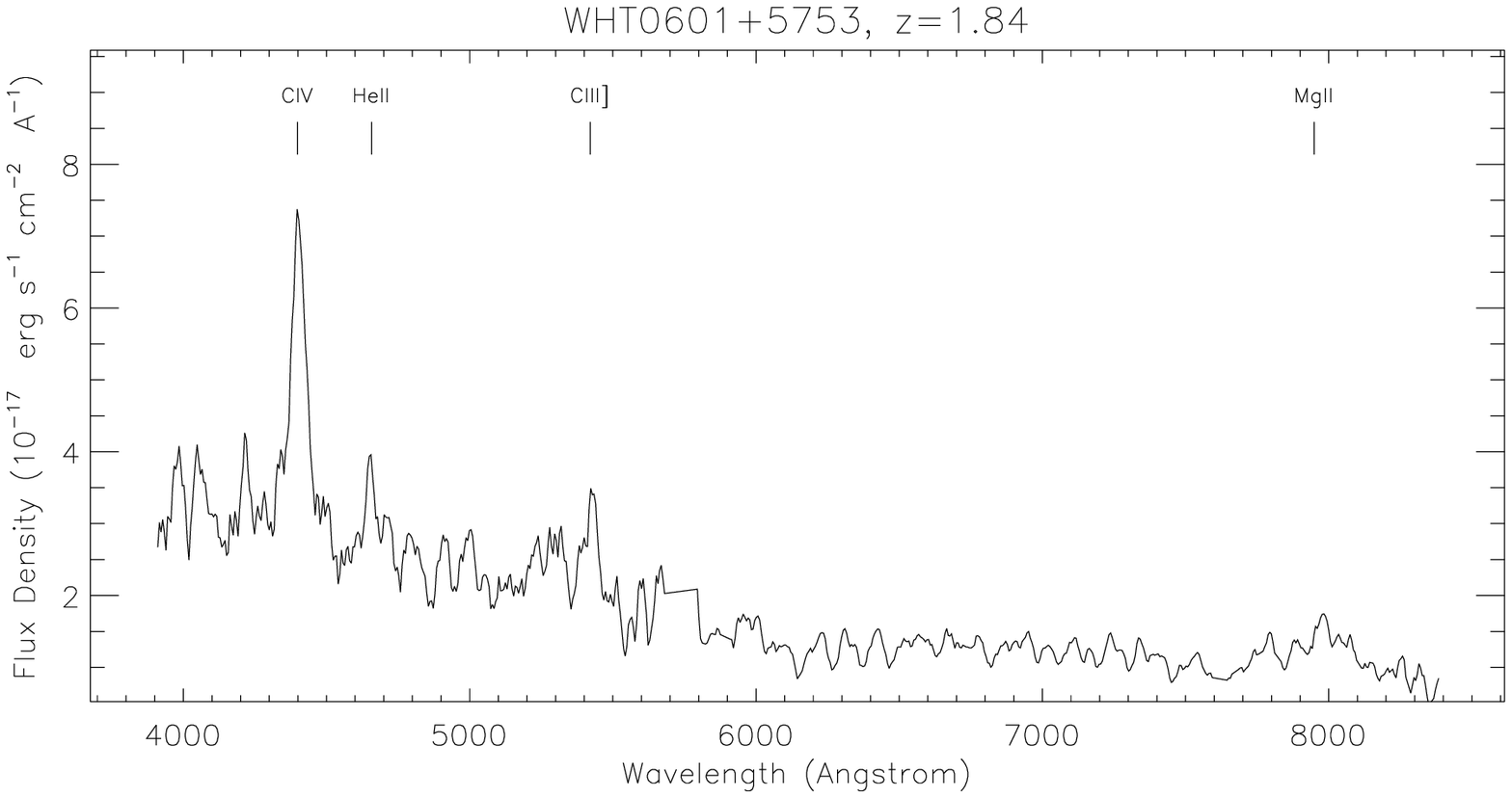,width=3.75cm,height=10.75cm,angle=90}}
\vspace{0.5cm}
\caption{\label{optspec}The optical spectra.}
\end{figure*}
\addtocounter{figure}{-1}
\begin{figure*}
\hbox{
\psfig{figure=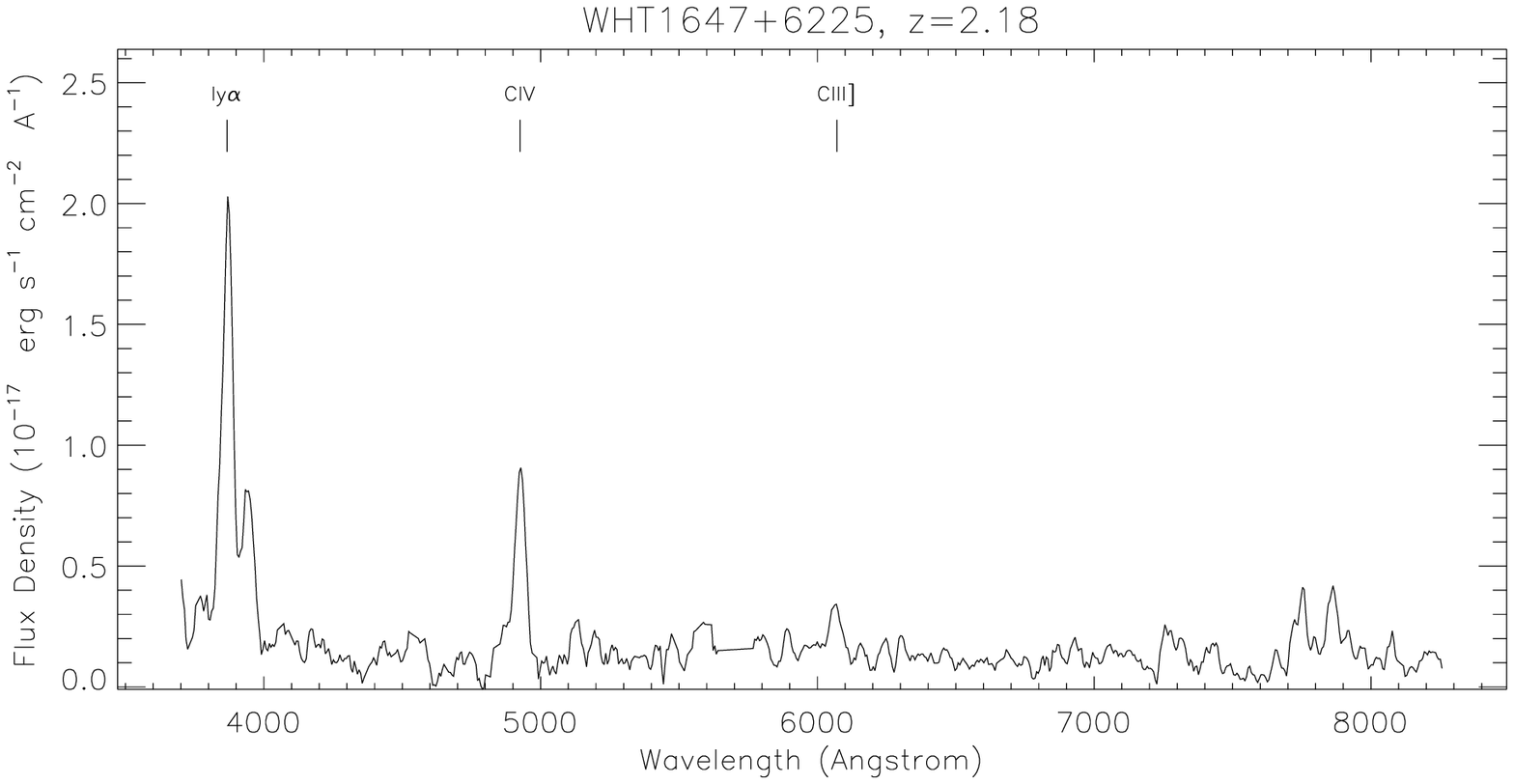,width=3.75cm,height=10.75cm,angle=90}
\psfig{figure=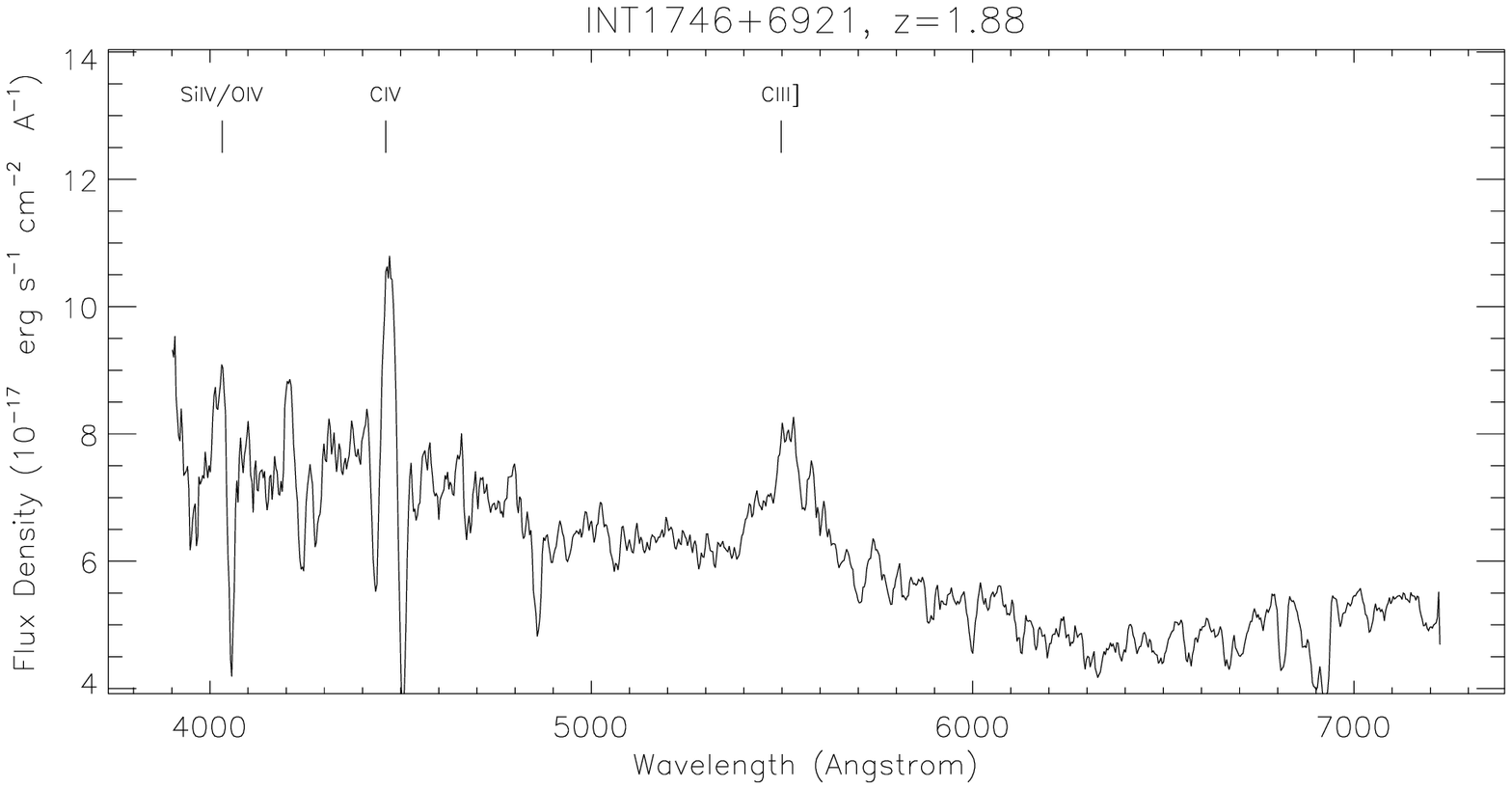,width=3.75cm,height=10.75cm,angle=90}
\psfig{figure=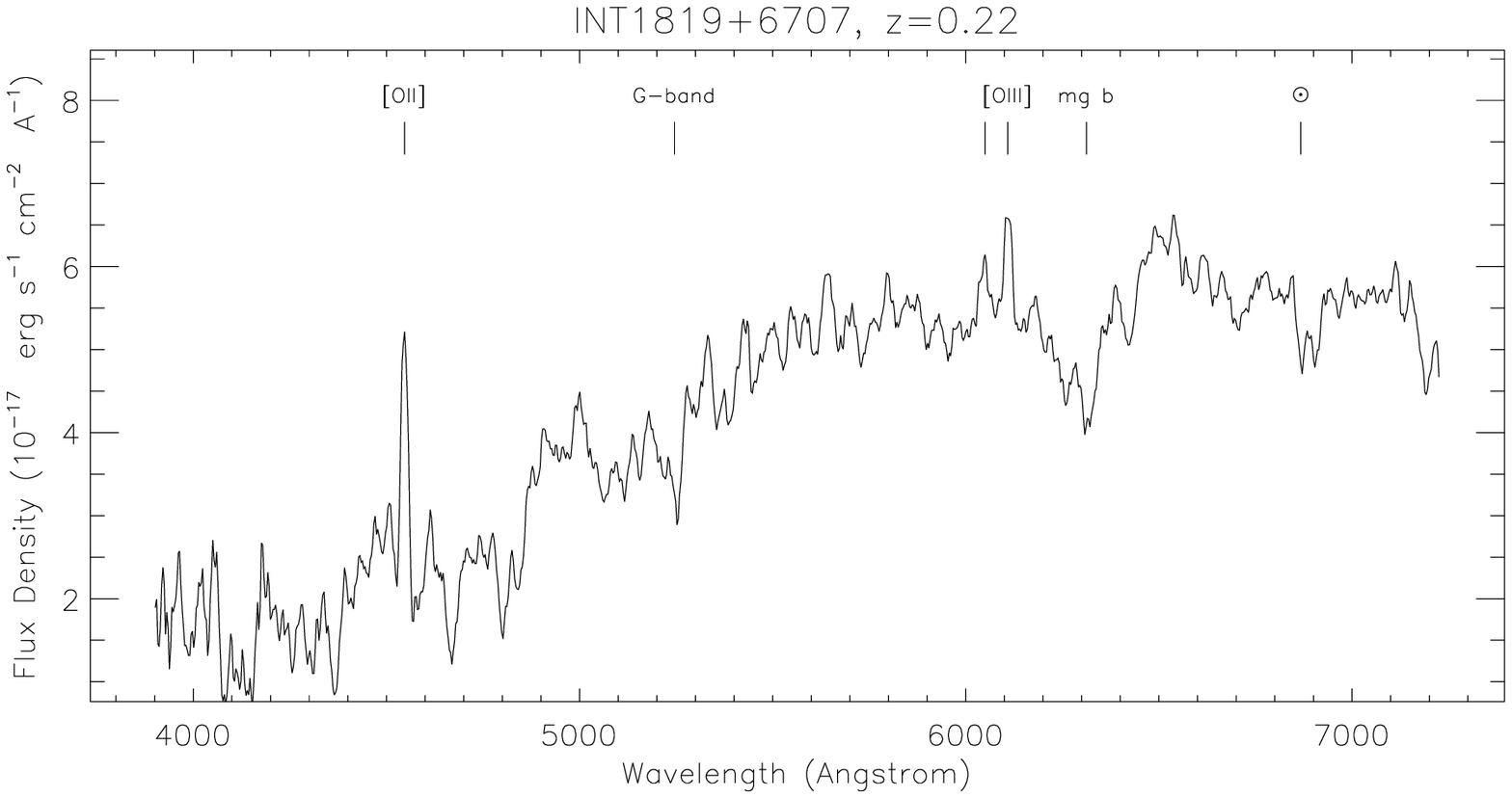,width=3.75cm,height=10.75cm,angle=90}
\psfig{figure=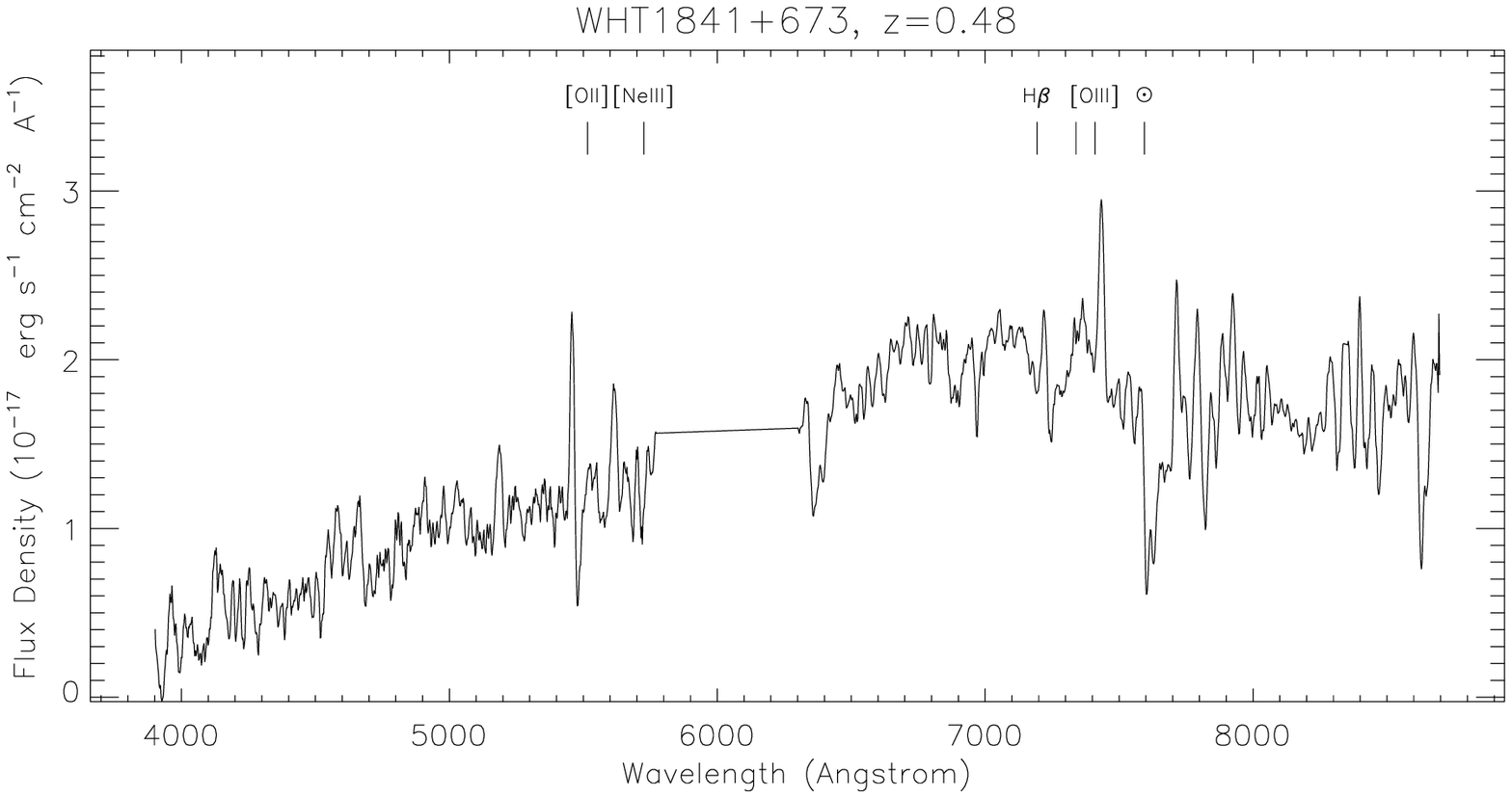,width=3.75cm,height=10.75cm,angle=90}}
\hbox{
\psfig{figure=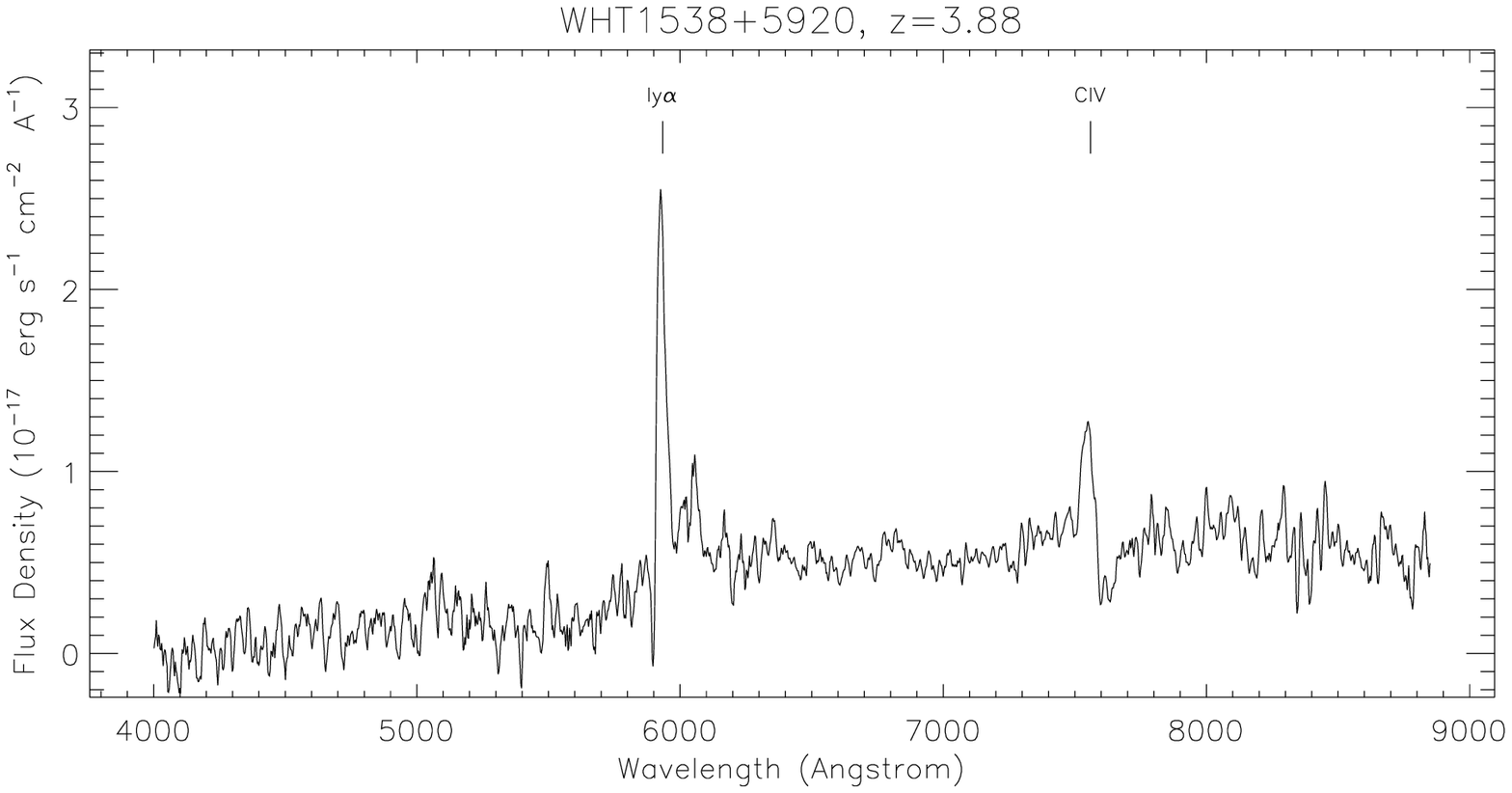,width=3.75cm,height=10.75cm,angle=90}
\psfig{figure=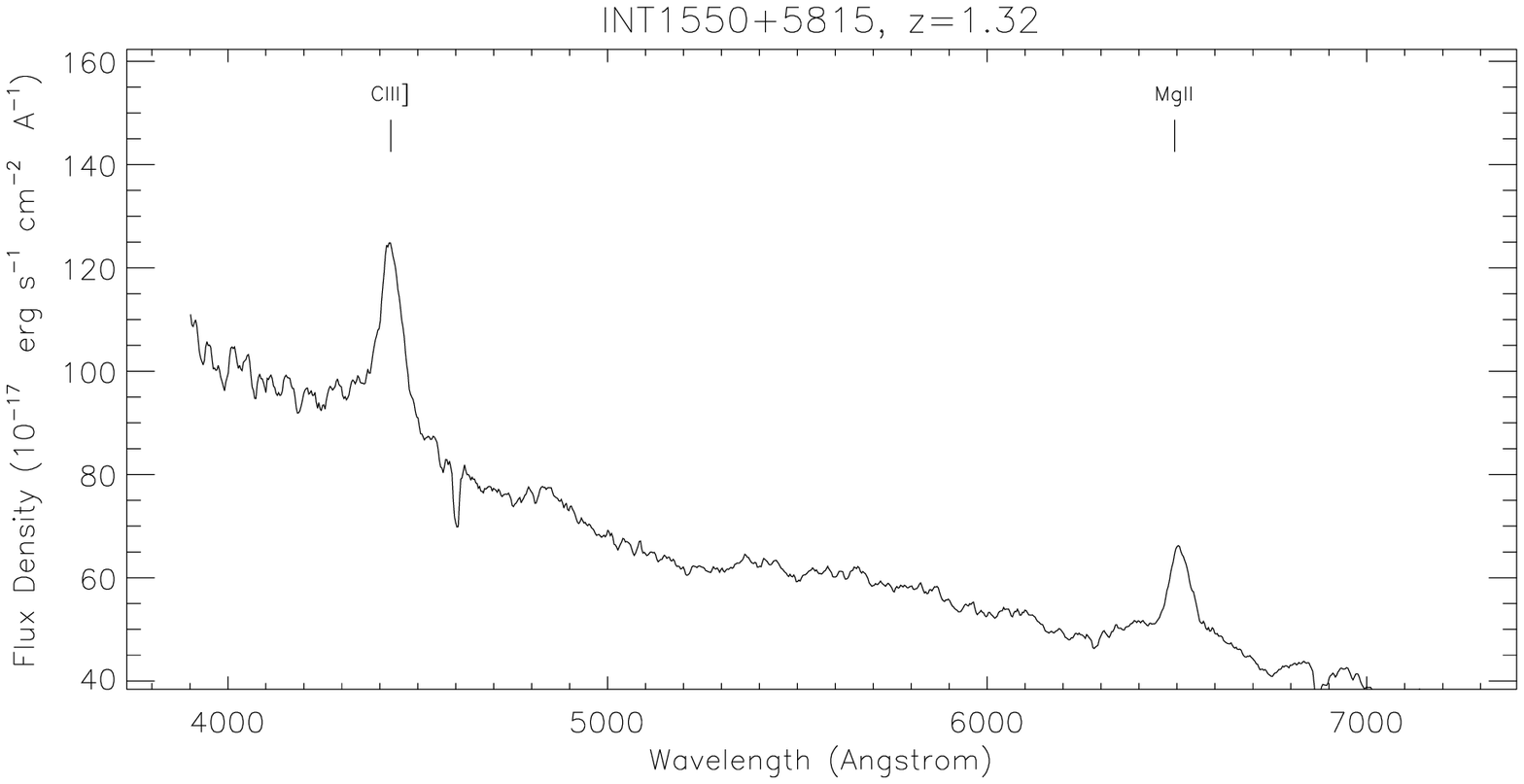,width=3.75cm,height=10.75cm,angle=90}
\psfig{figure=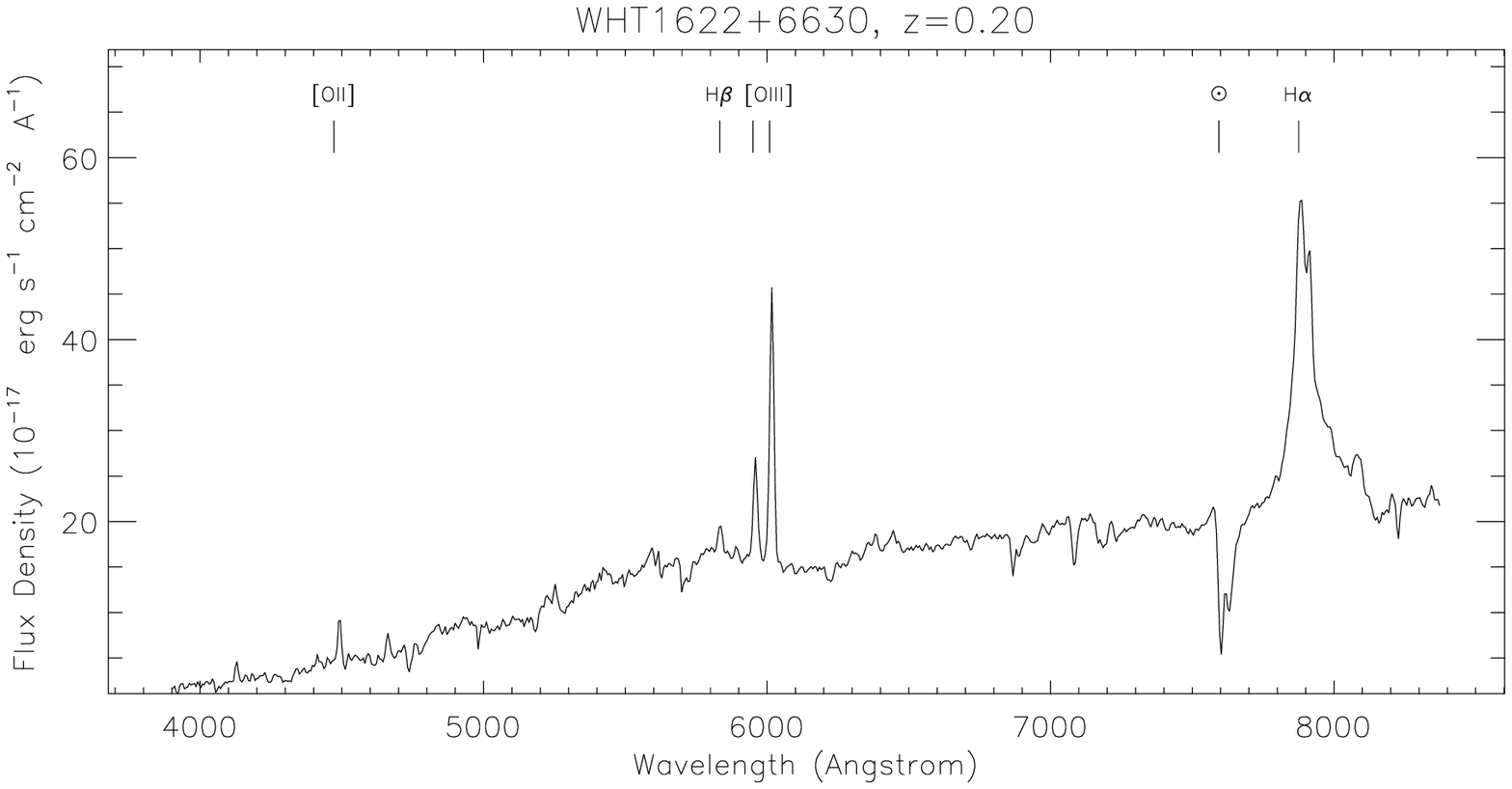,width=3.75cm,height=10.75cm,angle=90}
\psfig{figure=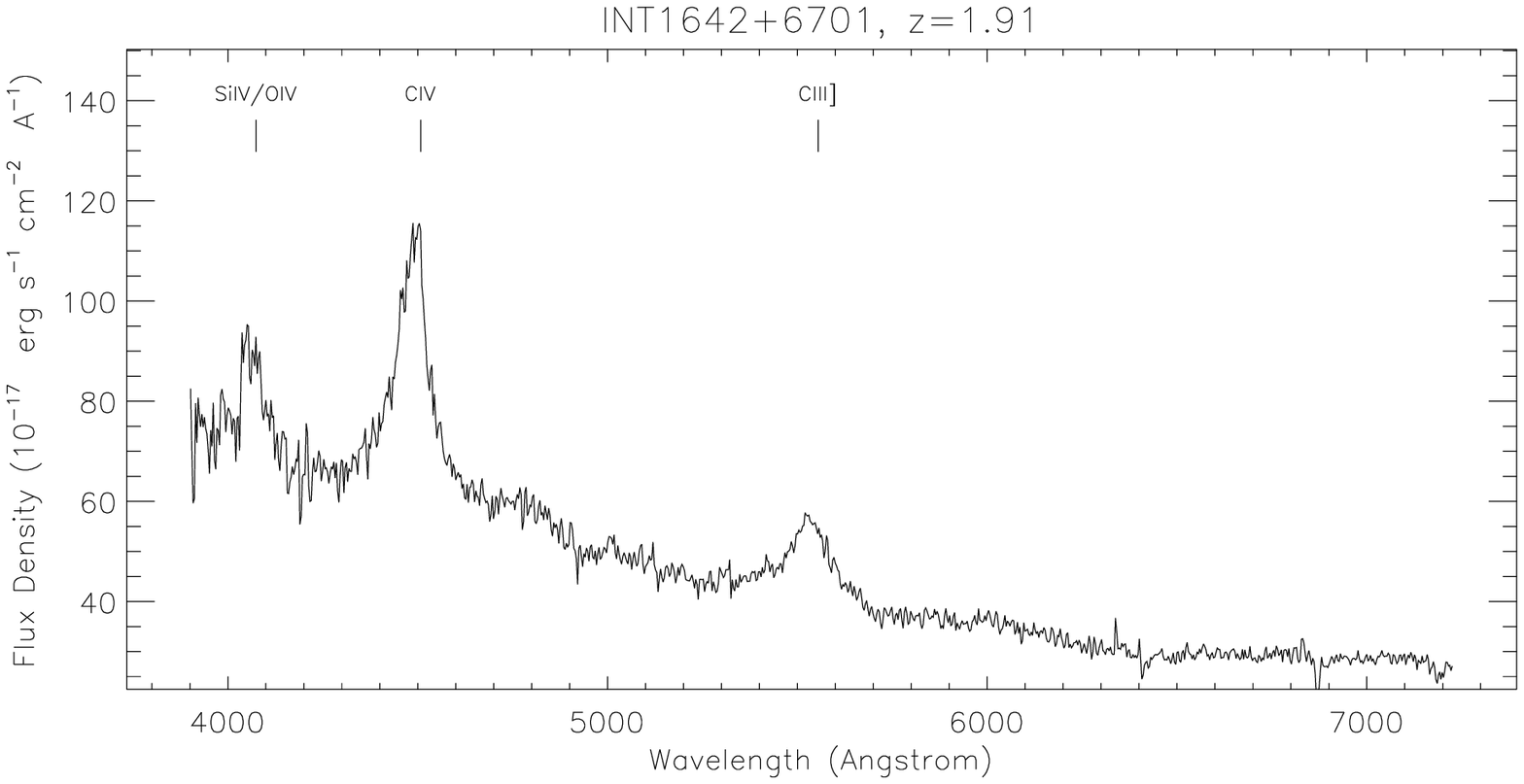,width=3.75cm,height=10.75cm,angle=90}}
\vspace{0.5cm}
\caption{Continued...}
\end{figure*}
\addtocounter{figure}{-1}
\begin{figure*}
\hbox{
\psfig{figure=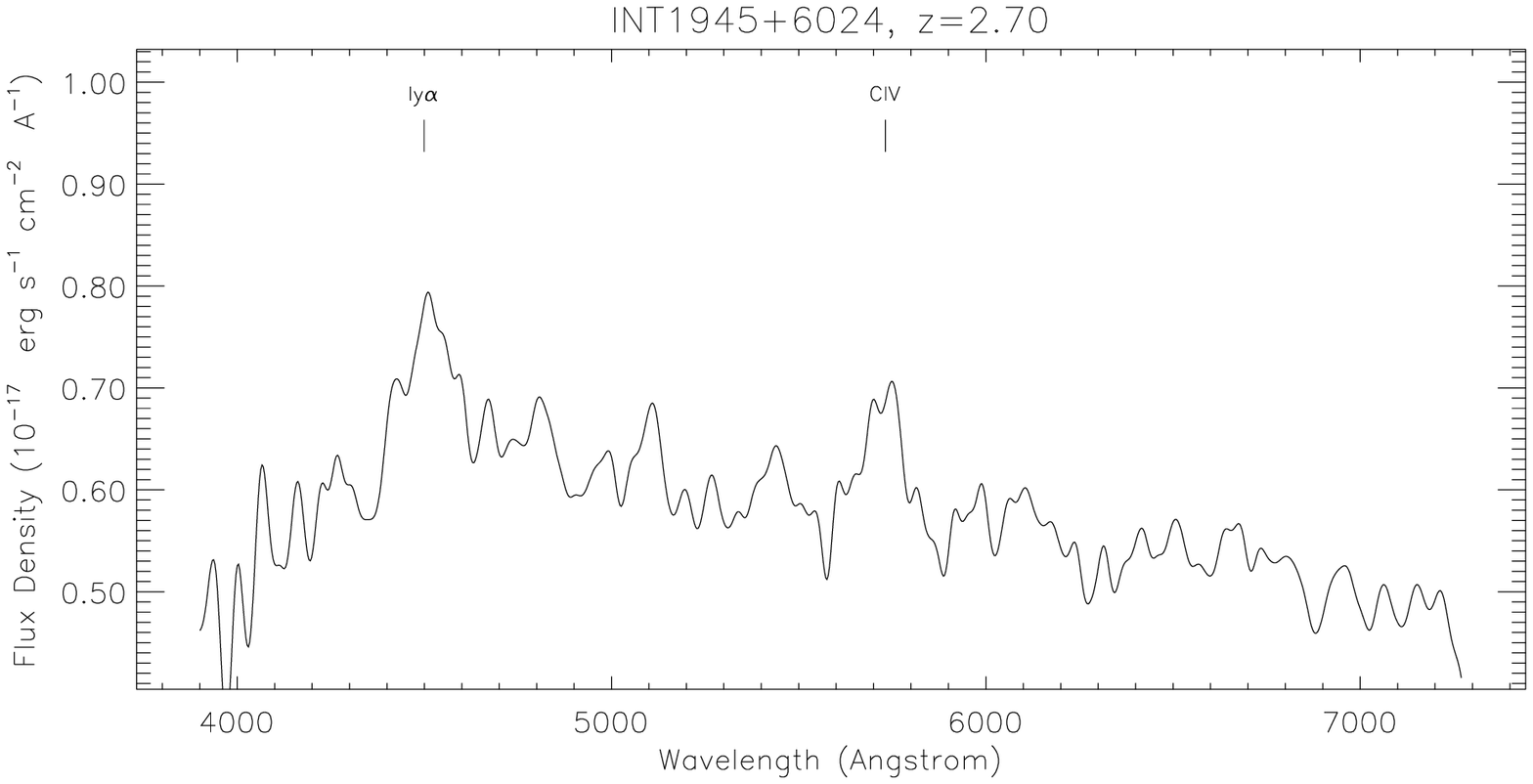,width=3.75cm,height=10.75cm,angle=90}
\psfig{figure=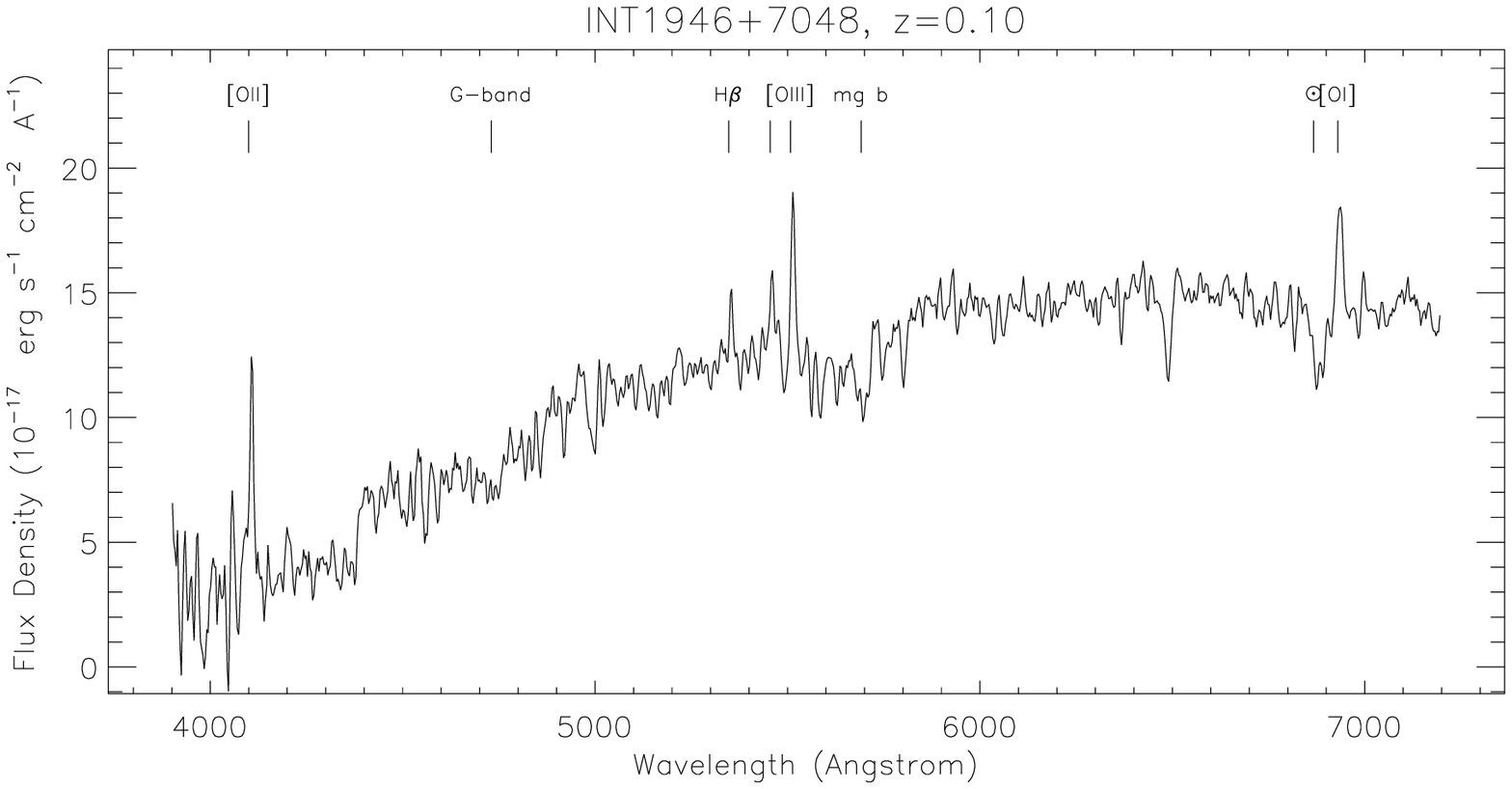,width=3.75cm,height=10.75cm,angle=90}
\psfig{figure=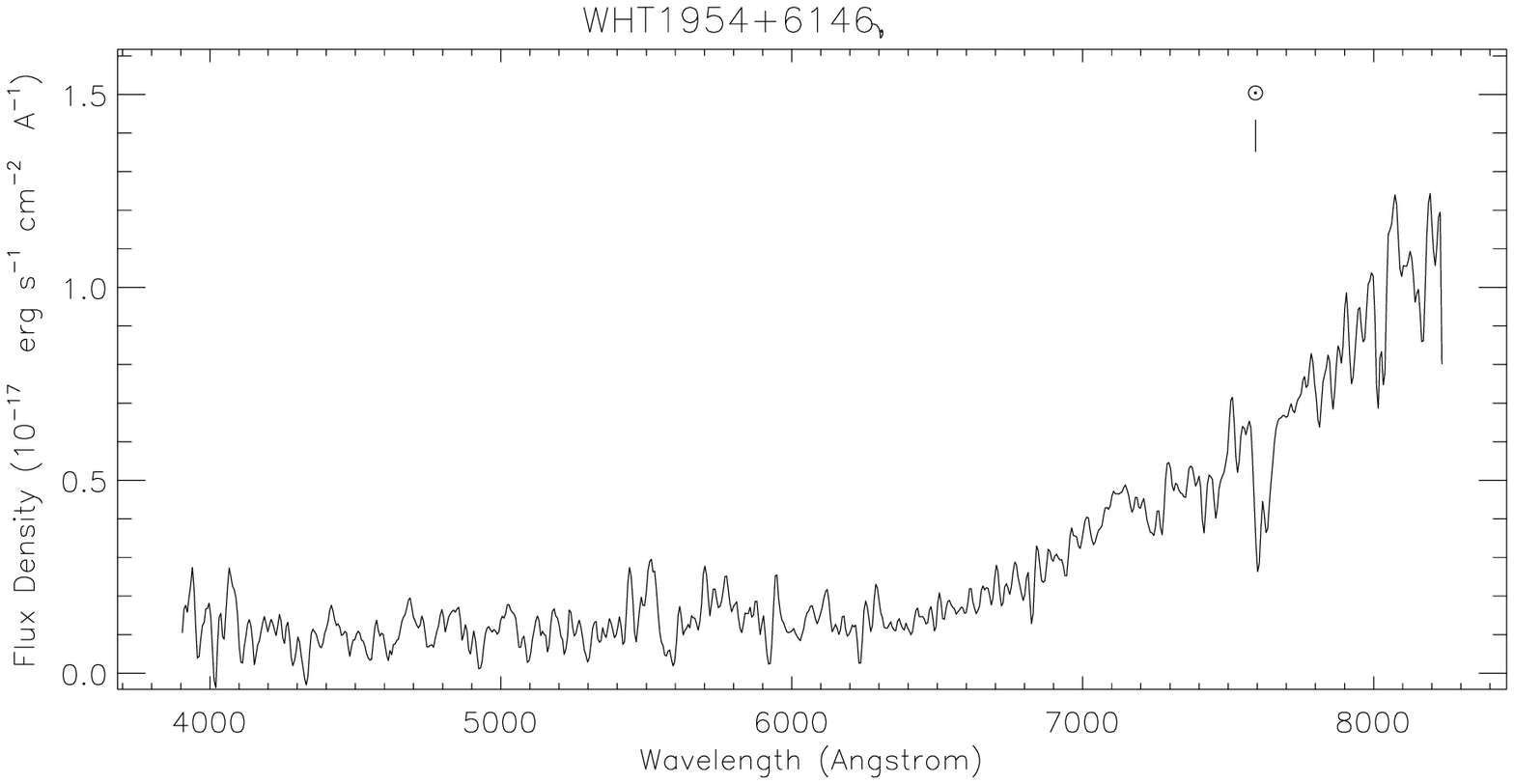,width=3.75cm,height=10.75cm,angle=90}
\psfig{figure=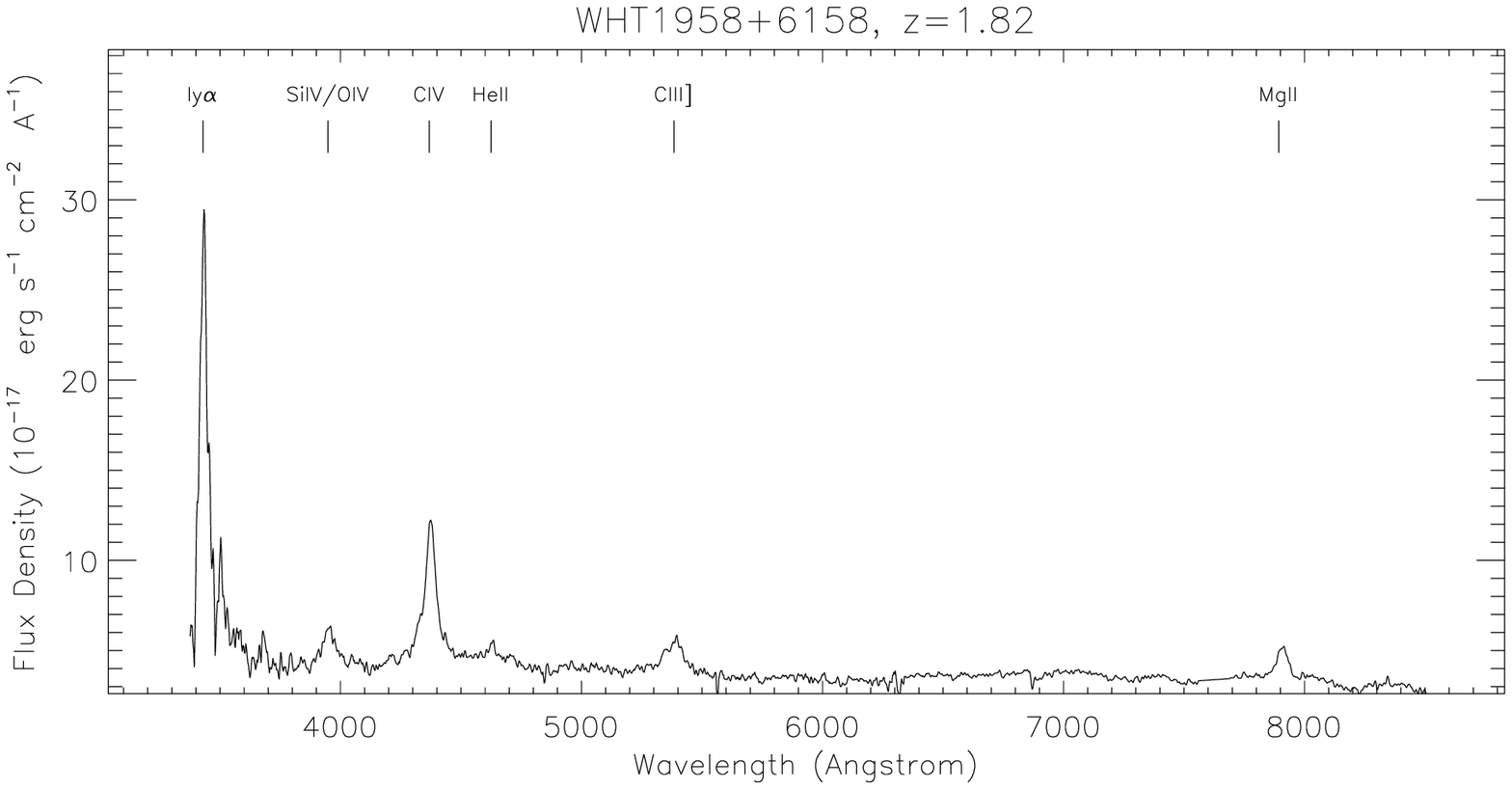,width=3.75cm,height=10.75cm,angle=90}}
\vspace{0.5cm}
\caption{Continued...}
\end{figure*}

\subsection{The Optical Spectra of GPS Galaxies}

\begin{figure*}
\centerline{
\psfig{figure=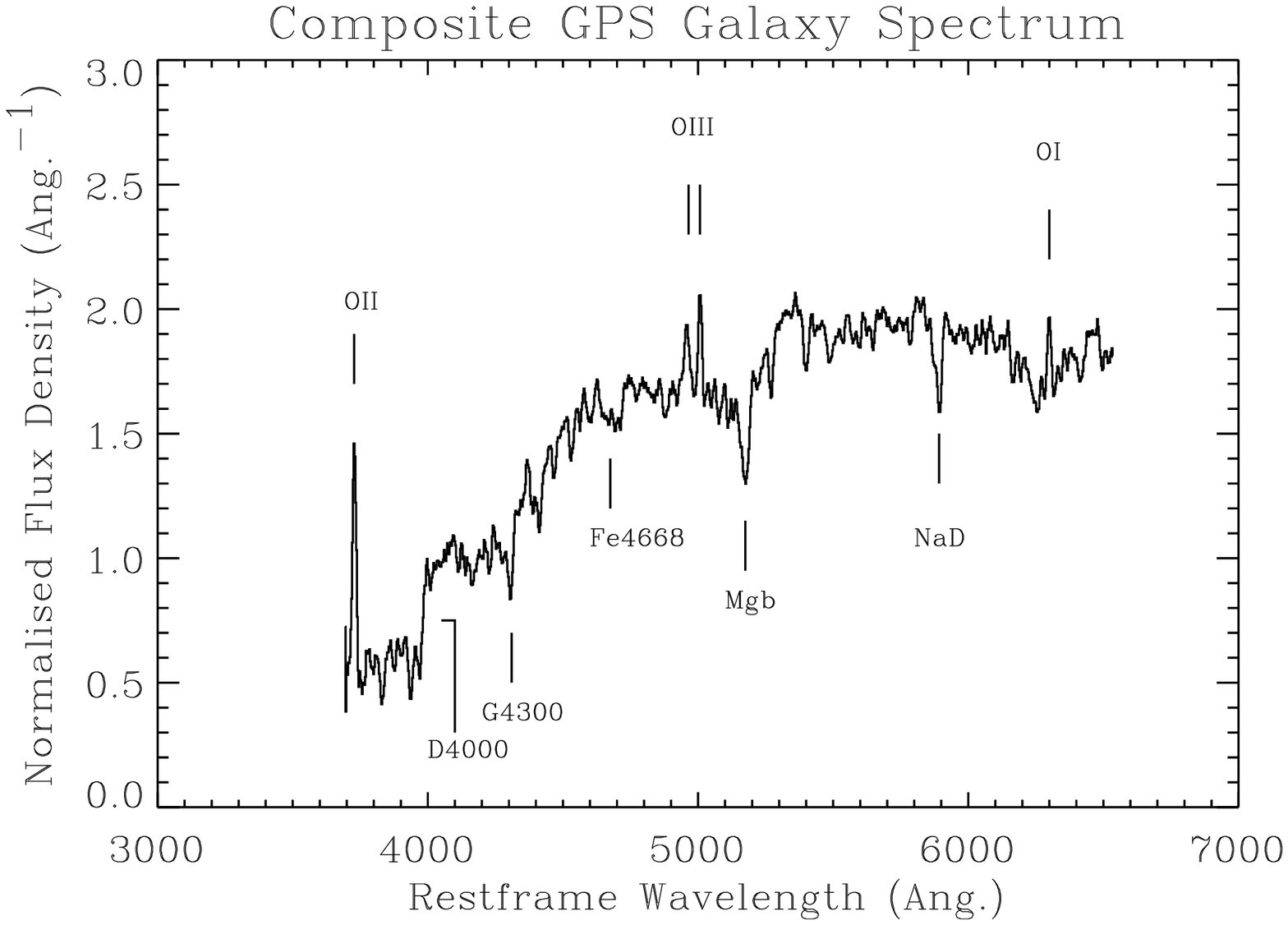,width=11.5cm}
\psfig{figure=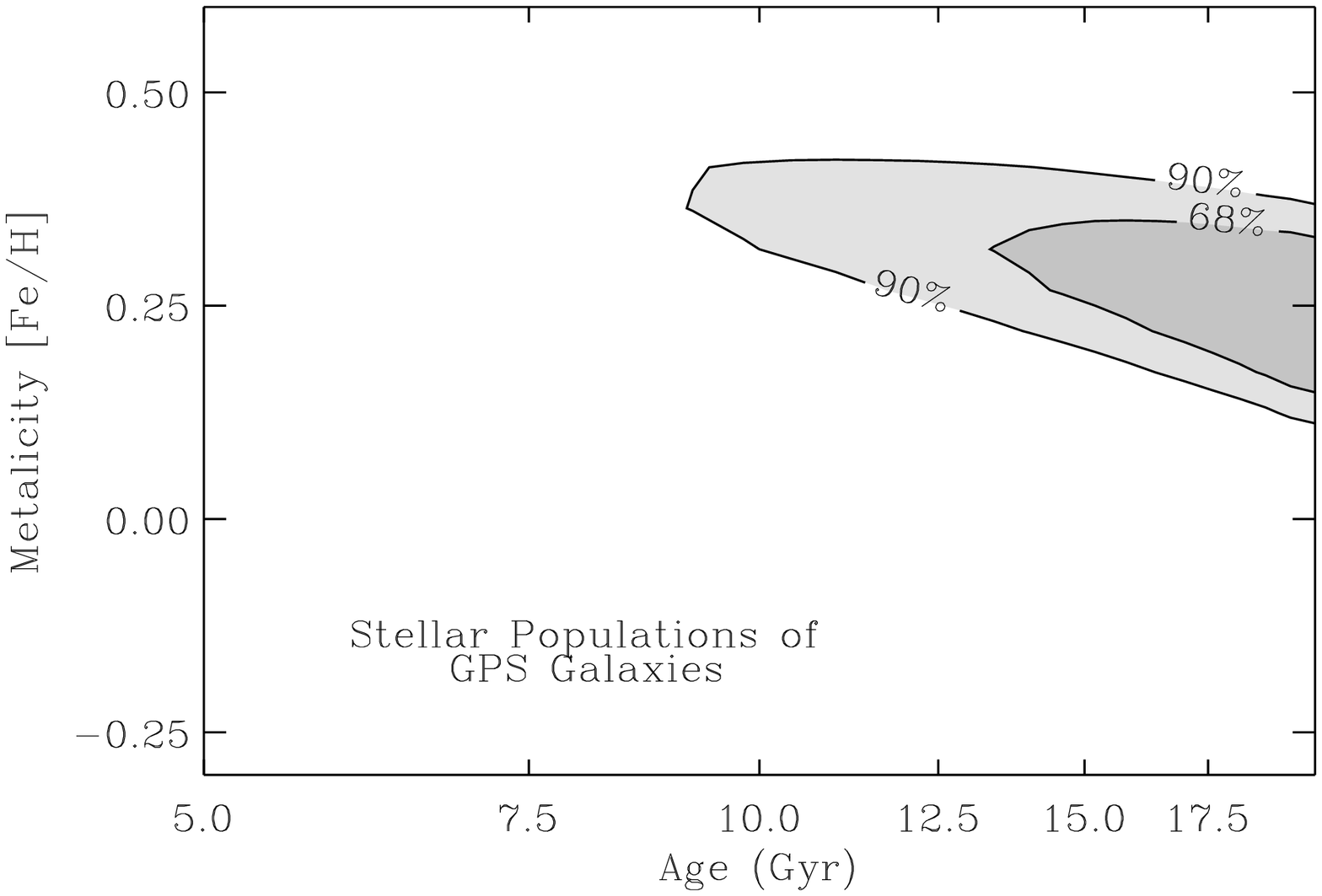,width=7cm,height=8cm}}
\caption{\label{compgal}(left) Composite GPS galaxy spectrum constructed
from the three low redshift galaxies in our sample 
(B0830+5813 at z=0.09, B1819+6707 at z=0.22, and B1946+7048 at z=0.10). 
Indicated are the 4000 \AA \ break
(D4000), the emission lines OI, OII and OIII, and the stellar absorption 
features Fe4668 \AA, the G-band at 4300 \AA \ (G4300), Mgb at 5170 \AA, and
NaD at 5900 \AA. (right) Confidence contours for the fit of the Worthey 
(1994) models to
the stellar absorption lines for a range of ages and metallicities. The 
absorption line strengths are compatible with high metallicity ($[$Fe/H$]>$0.2)
old ($>$9 Gyr) stellar populations.}
\end{figure*}

We obtained reliable spectra for six GPS 
galaxies, namely B0531+6121, B0830+5813, BG1622+6630, B1819+6707, 
B1841+6715 and B1946+7048. The spectra of two of the galaxies are clearly 
dominated 
by light coming from the active nucleus: B0531+6121 has a powerlaw spectrum
and can be classified as a narrow-line radio galaxy with bright Seyfert 2-like
emission lines. B1622+6630 can be classified as a Seyfert 1.5 with broad
(FWHM $\sim$3500 km/s) H$\alpha$ emission, and resembles the well known GPS 
galaxy B1404+286 (OQ208, MrK668, eg. O'Dea et al. 1991, 
and references therein). The broadline region of B1622+6630 seems to be 
heavily obscured and only visible in the red part of the spectrum. 
Unlike in B0531+6121, stellar absorption line features and the 4000 \AA$ $ 
break are visible in the blue part of the spectrum of B1622+6630 indicating 
that the non-stellar nuclear light is less prominent in the blue. 
The emission lines of the other four GPS galaxies are 
typical for radio galaxies, showing [OII] $\lambda 3727$, the [OIII] $\lambda
4959$ and $\lambda 5007$ lines, H$\alpha$/NII, H$\beta$ and in one case 
(B1946+7048) [OI] $\lambda 6300$. Their [OII]/[OIII] line ratios 
indicate low ionisation.

The optical to near-infrared broadband colours and magnitudes are consistent
with GPS galaxies being giant elliptical galaxies with old stellar populations
 (Snellen et al., 1996a,b.; O'Dea et al. 1996). The spectra obtained here can 
be used to determine whether the properties of the stellar absorption lines 
are consistent with this hypothesis. 
Stellar absorption lines are a more powerful tool  
than broadband colours, because the range of absorption features can be
used to disentangle age and metalicity effects (Worthey 1994). Furthermore,
the measured equivalent widths of the lines are less influenced by 
calibration errors. The three highest signal-to-noise spectra of the
most nearby GPS galaxies were shifted to the rest-frame and combined 
together to form a composite GPS galaxy spectrum (figure \ref{compgal}). 
The spectrum clearly shows the 4000 \AA \ break and several deep stellar 
absorption lines, with as most prominent, the G-band at 4300 \AA, Fe4668 \AA,
Mgb at 5175 \AA \  and NaD at 5900 \AA, with equivalent widths of 
6.7, 7.3, 5.0 and 4.5 \AA \ respectively. The absorption line indices were
compared with the Worthey (1994) models to estimate their age and metalicity.
According to these models, the dominant light of 
these galaxies is coming from a stellar population with an age $>9$ Gyr and
a metalicity between 1.5 and 2.5 times solar (figure \ref{compgal}).

Optically selected ellipticals show a strong correlation between 
absorption line strengths and mass (Dressler et al. 1987): 
the more massive the galaxy, the 
deeper the absorption lines. This indicates that giant ellipticals
are more metal rich and/or have older stellar populations than smaller 
ellipticals. The deep stellar absorption lines in the GPS galaxies
are therefore consistent with them being among the most massive
ellipticals, as found by Snellen et al. (1996a, 1996b). 
This makes it unlikely that a non-stellar contribution from
the active nucleus is present in the optical continuum of these galaxies. 
Even a small
contribution of, say 5\%, would make the intrinsic absorption lines
5\% deeper, resulting in even higher ($\sim$ 20\%) metalicities and/or ages,
which would correspond to unlikely ages higher than the age of the universe,
and masses larger than those for central cluster galaxies.
The claim that GPS galaxies are ideal objects to study the 
cosmological evolution of giant ellipticals seems therefore justified.

\subsection{The Optical Spectra of GPS Quasars}

In optical studies of quasars, GPS sources are generally treated as 
flat-spectrum radio sources. It is interesting to compare
the optical (restframe UV) spectra of GPS quasars with those of the general 
population of flat-spectrum quasars, to see whether a distinction can be
made between the two classes of objects. Any differences may provide 
important evidence on the influence Doppler boosting and orientation 
have on the optical and radio spectra of GPS quasars.

\subsubsection{The composite GPS quasar spectrum \label{compsect}}

Baker and Hunstead (1995) have studied a sample of flat spectrum quasars
as a function of their ratio of radio core-to-lobe flux density to reveal
the effects of orientation. They
 divided the quasars in their sample into three
sub-samples according to their radio core-to-lobe ratios $R$; $R>1$ (core
dominated), $0.1<R<1$ (intermediate), and $R<0.1$ (lobe dominated).
 They found that core-dominated quasars 
have bluer optical spectra, a stronger 3000 \AA \ bump and smaller equivalent
widths than lobe dominated quasars. All these trends can be accounted for
assuming an increase in dust extinction with viewing angle.
In addition, Baker and Hunstead (1995) found the spectra of CSS quasars 
to have evidence of even more reddening, with steep powerlaw continua,
strong low-ionisation narrow-line emission, and no 3000 \AA \ bump.

Following the same procedure as Baker and Hunstead (1995) we produced
a composite GPS quasar spectrum. To derive the composite spectrum, 
each individual spectrum was shifted to the quasar rest frame, and fitted 
with a power-law, $F(\nu) \propto \nu ^k$, in
the wavelength ranges $1300-1500$, $1700-1850$, and $1950-2200$ \AA , thereby
avoiding the influence of emission lines and contamination by the 3000 \AA \ 
bump. This power law fit was used to find the level of the continuum at
 3000 \AA \ and
to normalise the spectra at this wavelength. The redshifted and normalised
spectra were then co-added without using any weighting. Because of obvious
flux calibration problems, the data from the red ISIS arm of B0601+5753 and
B0758+5929, and the data blueward of 4500 \AA \  of B0758+5929 and B0826+7045,
could not be used in this process.

\begin{figure}
\centerline{
\psfig{figure=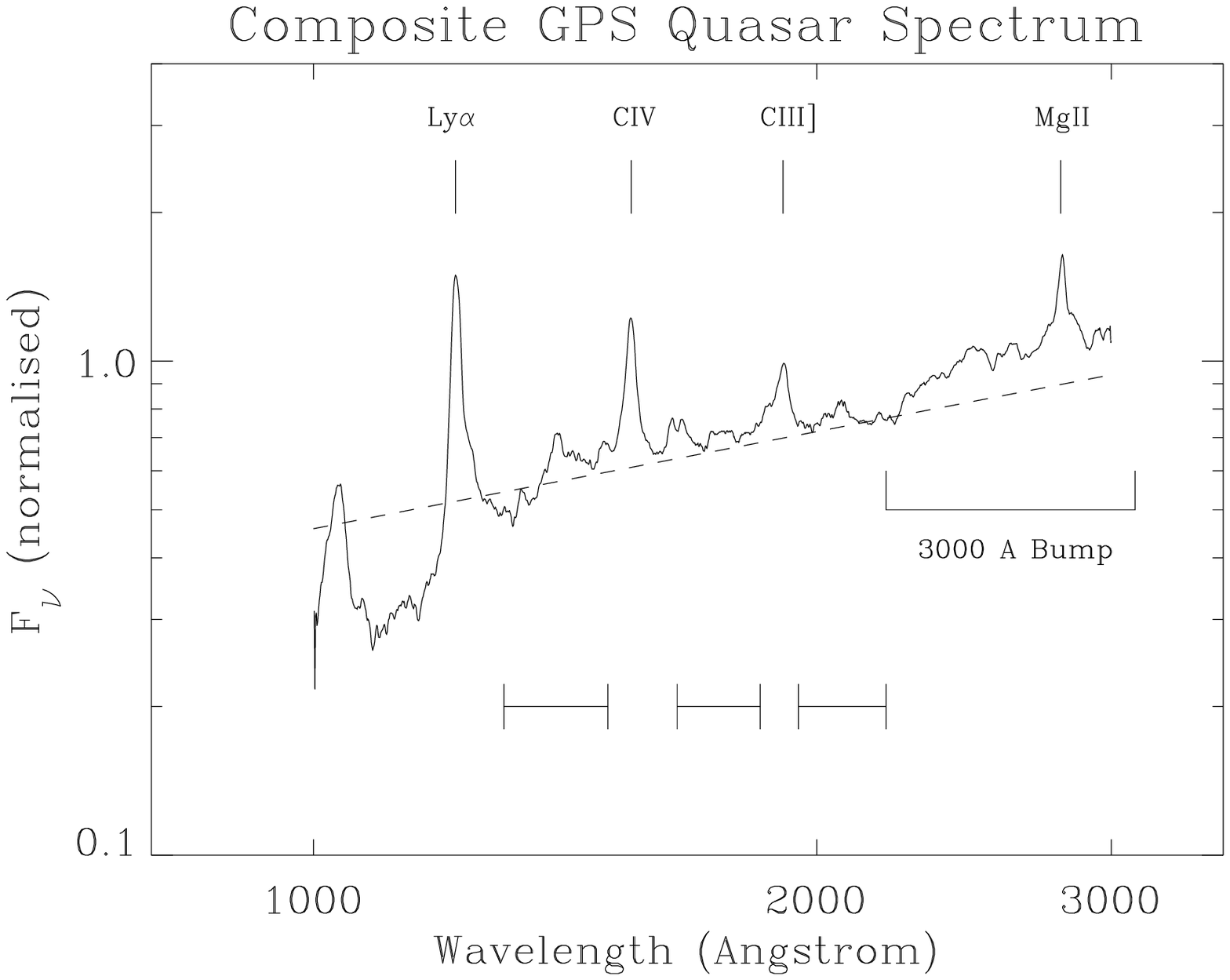,width=9cm}}
\caption{\label{composite}
A composite spectrum of the GPS quasars in our sample for which we have 
obtained spectra. The dashed line indicates the fit to the continuum between
1300 and 2200 \AA . Redward of 2300\AA\  the 3000\AA \ bump is visible.
The composite spectrum is similar to that of flat spectrum quasars 
(Baker and Hunstead 1995) and
more like core-dominated than lobe-dominated quasars,
but clearly different from that of CSS quasars.}
\end{figure}

By
normalising the spectra at a given wavelength, spectral slope characteristics
have largely been preserved. However, to be able to make a comparison with the
Baker and Hunstead results, the spectra have been normalised at the same
wavelength (3000 \AA), which for all our quasars is at or even beyond the red
end of their observed spectrum. This makes the contribution of the blue
quasars stronger than the red quasars in the composite spectrum, but this
effect should be similar in the Baker and Hunstead results in this part of the
spectrum. The composite spectrum is shown in figure \ref{composite}.
Below 1500 \AA \ the composite spectrum is very inaccurate, because
only a few quasars, the ones with the highest redshifts, contribute to this
part of the spectrum. The fitted power law, indicated by the dashed line, 
has a spectral index of 0.8, using the wavelength ranges $1300-1500$, 
$1700-1850$, and $1950-2200$ \AA. However, when using only the wavelength 
ranges $1700-1850$, and $1950-2200$ \AA, a spectral index of 0.7 is obtained, 
indicating the uncertainty in the fit caused by the spectral curvature. 
Even taking into account this uncertainty the 3000 \AA \ bump is clearly 
visible. The equivalent widths of Ly$_{\alpha}$, CIV, CIII$]$ and MgII are 
65 \AA , 25 \AA, 20 \AA \ and 25 \AA , respectively. 

The composite GPS quasar spectrum is clearly different from that of
CSS quasars and is similar to that constructed for
flat-spectrum quasars (Baker and Hunstead, 1995).  The
spectral slope measured for the composite GPS quasar spectrum is
comparable to that for the intermediate sub-sample (0.7), and the strength
of the 3000 \AA \ bump is consistent with both the core-dominated and
intermediate sub-sample. The equivalent widths of the mayor emission
lines matches better those of the core-dominated quasars
(Ly$_{\alpha}$ = 70 \AA, CIV = 50 \AA, CIII$]$ = 34 \AA and MgII = 43
\AA) than those of the lobe-dominated quasars (CIV = 120 \AA, CIII$]$ =
16 \AA \ and MgII = 71 \AA). The composite GPS quasar spectrum shows
that the average optical properties of GPS quasars are
indistinguishable from those of flat-spectrum quasars in general, but
that they appear to be more like core-dominated quasars than
lobe-dominated quasars.  It is likely that GPS quasars are also
present in the Baker and Hunstead (1995) sample. They will appear in
the core-dominated sub-sample because the $R$ values are measured at
$\sim 1''$ resolution. The composite GPS quasar spectrum is therefore
consistent with the Baker and Hunstead result. Note however that the redshift
distribution for the GPS quasars is different from that of
flat-spectrum quasars and that of CSS quasars.  For the samples used,
GPS quasars are biased towards the highest redshifts and CSS quasars
are biased towards the lowest redshifts, and within the sample of
flat-spectrum quasars, the core-dominated quasars are biased towards
higher redshifts than the lobe-dominated quasars. The cosmological
evolution of dust in the vicinity of quasars may therefore influence
this result, and any conclusions must be drawn with caution.
The evidence for a low dust extinction in GPS quasars may indeed indicate a 
small viewing angle as proposed by Baker and Hunstead for core-dominated
quasars, but it can also reflect a decrease in dust-content with redshift.

\subsubsection{The Baldwin effect in GPS quasars \label{sectbaldwin}}

A major effect in quasar spectra is the Baldwin effect, in which the 
equivalent width of the 
CIV $\lambda$1549 line decreases with increasing continuum luminosity, 
particularly 
in quasars that are flat spectrum radio sources (Baldwin 1977, Baldwin,
Wampler and Gaskell 1989). In photo-ionisation models for quasars, 
the line luminosity is linearly dependent on the continuum luminosity, but
is also a strong function of the density and ionisation parameters.
The Baldwin effect is believed to be a consequence of an 
ionisation parameter decreasing with luminosity 
(Mushotzky and Ferland, 1984). Interestingly, due to the fact that
the line flux is not Doppler boosted, the scatter in the CIV-continuum
luminosity relation can be used to put an upper limit on the average Doppler
factors for the optical continuum in GPS quasars (Kinney et al. 1985).

\begin{figure}
\psfig{figure=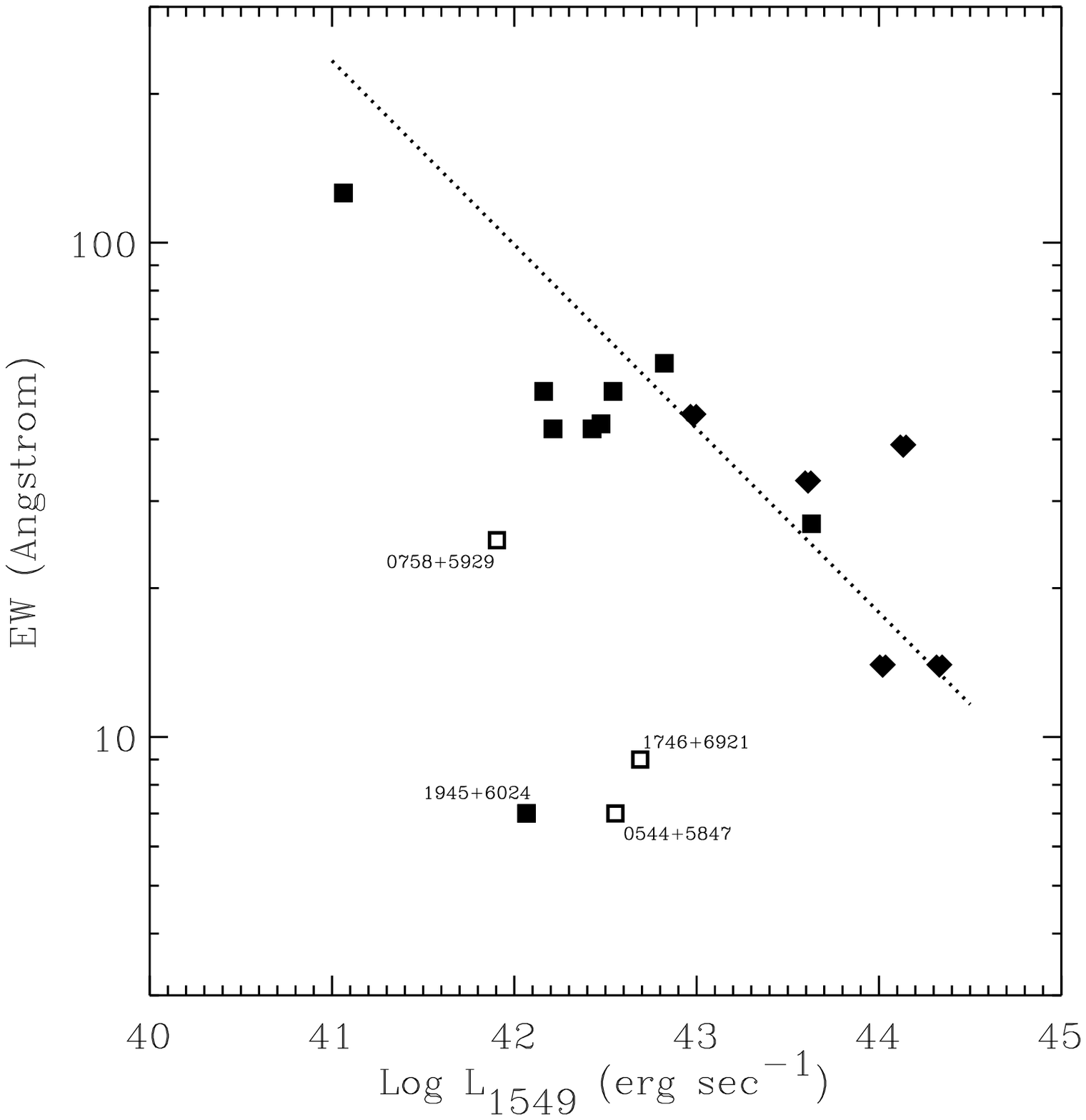,width=9cm}
\caption{\label{baldwin} The CIV Baldwin effect for GPS quasars. The squares 
indicate objects from the faint WENSS sample and the diamond symbols 
quasars from the Stanghellini et al. (1998) sample. The open squares indicate
quasars which show clear evidence for associated CIV absorption, which 
significantly reduces the observed CIV equivalent widths for these objects.
The quasar B1945+6024 is possibly affected by absorption as well, but 
the spectrum is of too low signal to noise to confirm this.
The dotted line shows the strength of the Baldwin effect in flat spectrum
quasars from Baldwin, Wampler and Gaskell (1989).}
\end{figure}

\begin{figure}
\psfig{figure=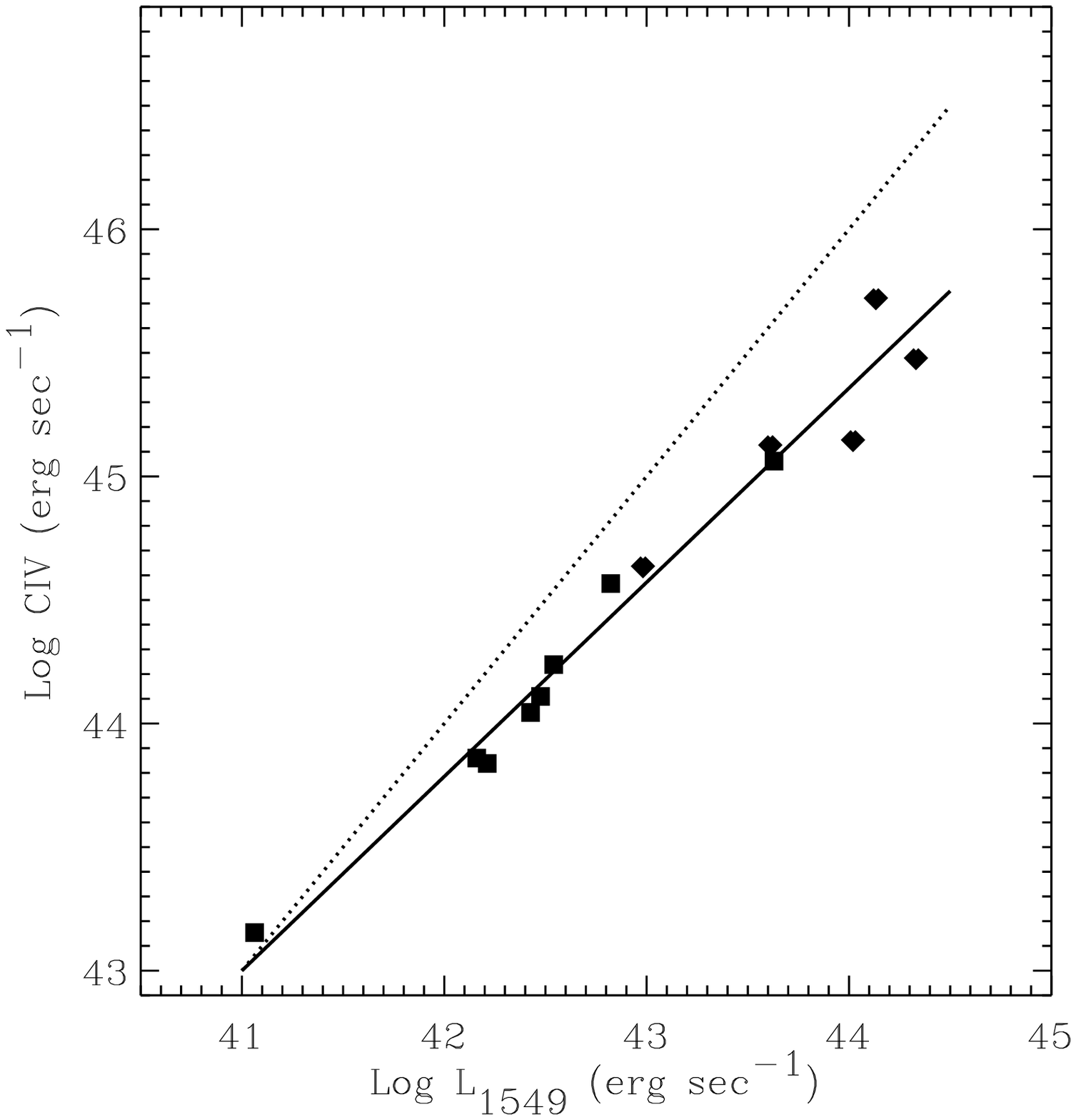,width=9cm}
\caption{\label{baldwin1} The continuum luminosity at 1549 \AA \ versus
the CIV line luminosity for GPS quasars, excluding the ones influenced 
by associated absorption. The deviation from a linear
correlation, indicated by the dotted line, is an alternative way of 
showing the Baldwin effect. The low dispersion in the relation between
the isotropic CIV line luminosity and the continuum luminosity indicates
that the continuum emission can only be mildly Doppler boosted.}
\end{figure}

To investigate the Baldwin effect in GPS quasars, our sample has been 
complemented
with objects from the complete sample of radio-bright GPS sources of 
Stanghellini et al. 
(1998). This results in an additional 5 GPS quasars at $z>1.9$ with 
their CIV emission-line subsequently redshifted to optical wavelengths.
The CIV emission line properties of these radio-bright GPS quasars 
are taken from the literature and shown in table \ref{brightCIV}. 
The Baldwin effect for the 17 GPS quasars is shown in figure \ref{baldwin}.
The spectra of 3 quasars show clear evidence that their CIV emission lines
are severely disrupted by associated absorption, which greatly 
decreases their line luminosities and equivalent widths. 
These objects, B0554+5847, B0758+5929, and B1746+6921, are indicated by open 
squares in figure \ref{baldwin}. The spectrum of quasar B1945+6024 
is of insufficient signal-to-noise to determine whether CIV absorption is present. However,
the low equivalent width with respect to the continuum luminosity indicates
that CIV absorption is indeed important.
Although evidence of the Baldwin effect for the remaining quasars is mainly
based on the one low luminosity source, its strength and scatter is 
comparable to that for flat spectrum quasars (dotted line, 
Baldwin, Wampler and Gaskell, 1989).
The continuum luminosity at 1549 \AA \ is plotted against the CIV line
luminosity in figure \ref{baldwin1}, with the solid line indicating
the best fit. A linear relation between the two quantities is indicated by 
the dotted line, and the difference between this linear relation and the 
observed relation is an alternative way of plotting the Baldwin effect.

The scatter in the CIV - continuum luminosity relation is only 30\%. Although
in general GPS quasar spectra show evidence for low dust contents,
the scatter may be partially induced by the ratio of  dust 
extinction towards the CIV and continuum emission regions changing from
quasar to quasar. More interestingly, the fact that 
the CIV emission line is emitted isotropically (and not too much affected 
by dust at small viewing angles to the line of sight) allows the scatter to be
used to put an upper limit on the optical Doppler boosting. 
For example, if the optical continuum is Doppler boosted, the emission
line luminosity is independent of viewing angle (disregarding extinction),
but the continuum luminosity is a strong function of viewing angle. 
Therefore a sample of quasars with viewing angles randomly distributed 
within a certain range will result in a larger scatter in
the CIV - continuum luminosity relation. A population of 
quasars with randomly distributed viewing angles within $45^{\circ}$
and with velocities of $\beta = v/c = 0.5$ have
optical continua which are on average Doppler boosted by a factor 5, and 
produce a 30\% scatter in the CIV - continuum luminosity relation. If a 
smaller  quasar opening angle is used, the scatter is lower due to a smaller 
range in viewing angles. However, if the quasar-opening-angle is reduced in
such a way that the scatter allows $\beta=0.9$, 
quasar-to-quasar variations in $\beta$ are likely to produce a much larger
scatter in the observed CIV-continuum luminosity relation.
The low scatter in the Baldwin effect therefore indicates that
the optical emission of GPS quasars is only mildly Doppler boosted, unless
there is only a small range in both $\beta$s and viewing angles.

\begin{figure}
\centerline{
\psfig{figure=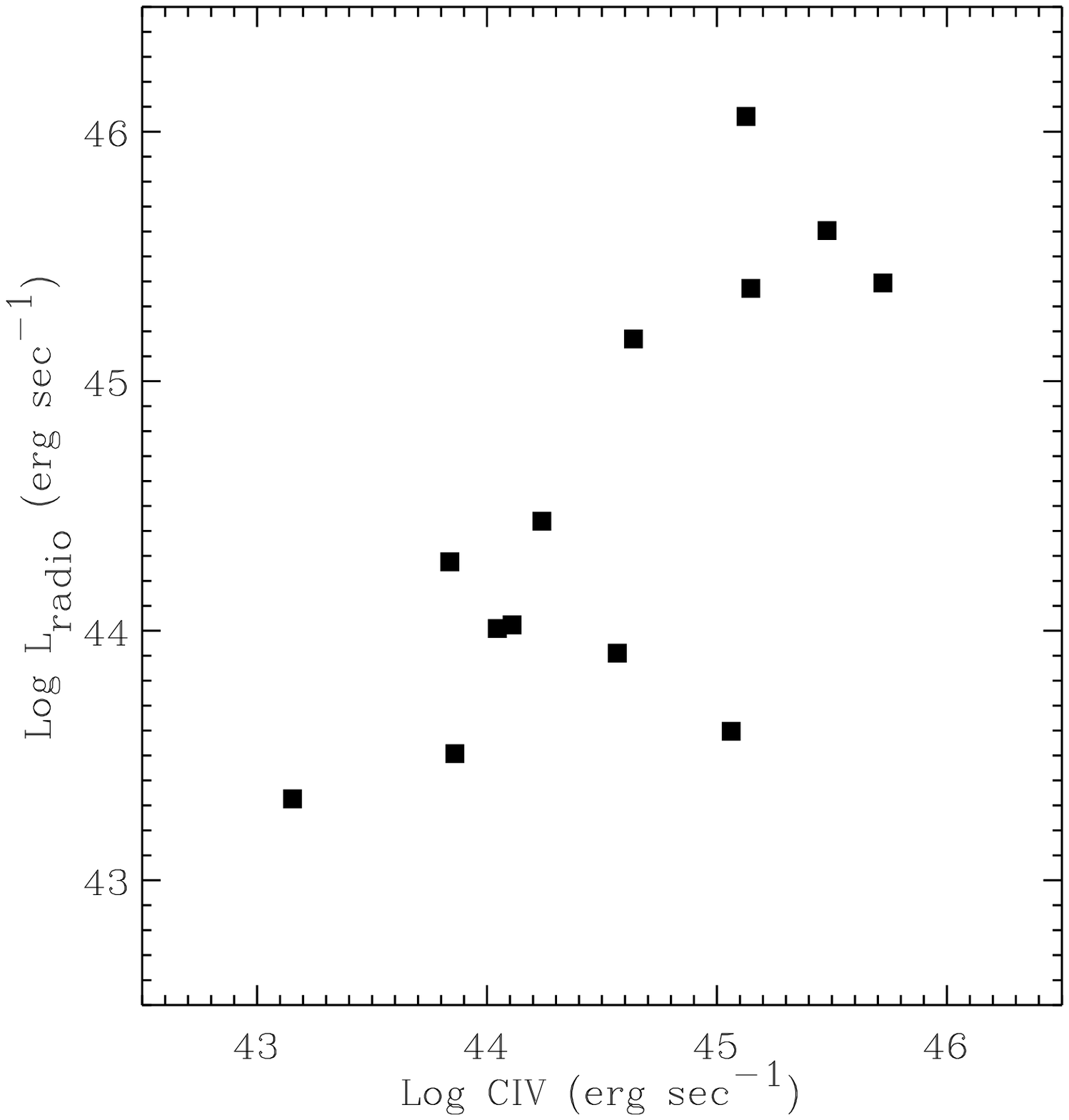,width=9cm}}
\caption{\label{contrad} The isotropic CIV line luminosity versus
the radio luminosity. Although the scatter is large in this  correlation, it provides an upper limit to the effect of Doppler boosting on the 
dominant radio components in GPS quasars.}
\end{figure}

A weak correlation between the CIV luminosity and radio luminosity 
(figure \ref{contrad}) enables us to perform a similar analysis on the radio 
emission. Assuming that the factor 3 scatter is produced by
source to source variations in Doppler boosting leads to an upper limit
of $\beta <0.85$ and the conclusion that the dominant radio emission in GPS 
quasars is on average not more Doppler boosted than by a factor of 15.

\begin{table}
\begin{tabular}{cccccr}\hline
Source&z&$S_{cont}$&$L_{CIV}$&W(CIV)&Ref.\\ 
      & & ($\frac{erg}{cm^2s\AA}$)    & ($\frac{erg}{cm^2s}$)   & (\AA)&\\ \hline
0237-233&2.223&1.2e-15&1.5e-13&39&1,2\\
0457+024&2.384&6.9e-17&1.1e-14&45&3\\
1442+101&3.544&2.3e-16&1.4e-14&14&4\\
2126-158&3.270&6.0e-16&3.6e-14&14&4\\
2134+004&1.936&5.4e-16&5.2e-14&33&3\\ \hline
\end{tabular}

References:\\
1: Wills et al. 1993\\
2: Wilkes et al. 1983\\
3: Baldwin, Wampler and Gaskell, 1989\\
4: Osmer, Porter and Green, 1994\\
\caption{\label{brightCIV} The GPS quasars from the radio-bright GPS sample
of Stanghellini et al. (1998) with measured CIV $\lambda$1549 \AA \ emission
line properties. Column 1 gives the name, column 2 the redshift, column 3
the continuum luminosity at 1500 \AA, column 4 the line luminosity, column 5 the restframe
equivalent width, and column 6 the references.}
\end{table}

\subsection{The Redshift Distributions of GPS Galaxies and Quasars}

\begin{figure}
\centerline{
\psfig{figure=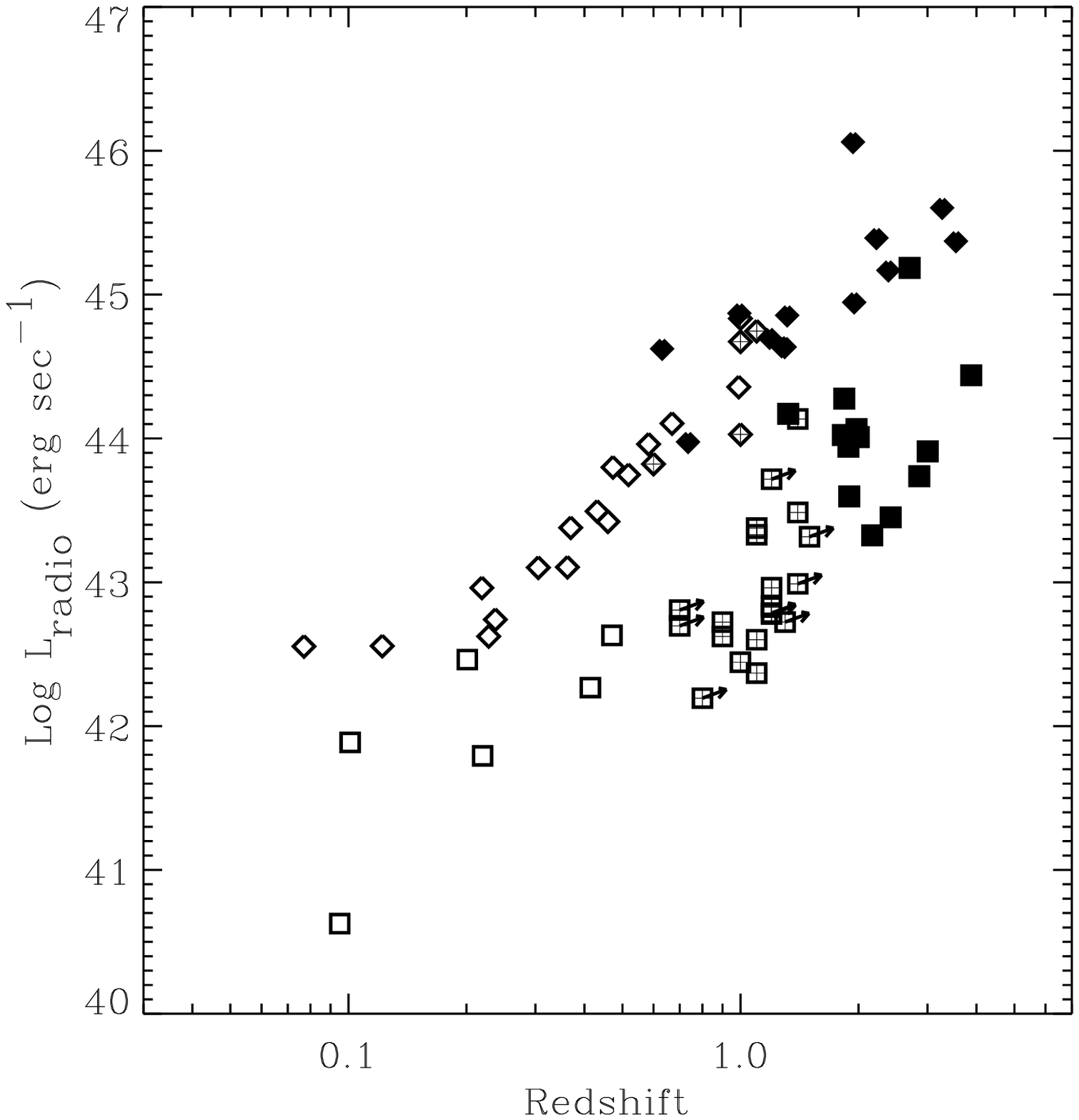,width=9cm}}
\caption{\label{zradio}The redshift versus radio luminosity for the GPS sources
in the radio-bright sample of Stanghellini et al (1998, diamonds) and our
radio-faint sample (squares). The open symbols indicate galaxies, and the 
filled symbols quasars. The crossed squares are galaxies with redshifts 
estimated from their optical apparent.}
\end{figure}

\begin{figure}
\centerline{
\psfig{figure=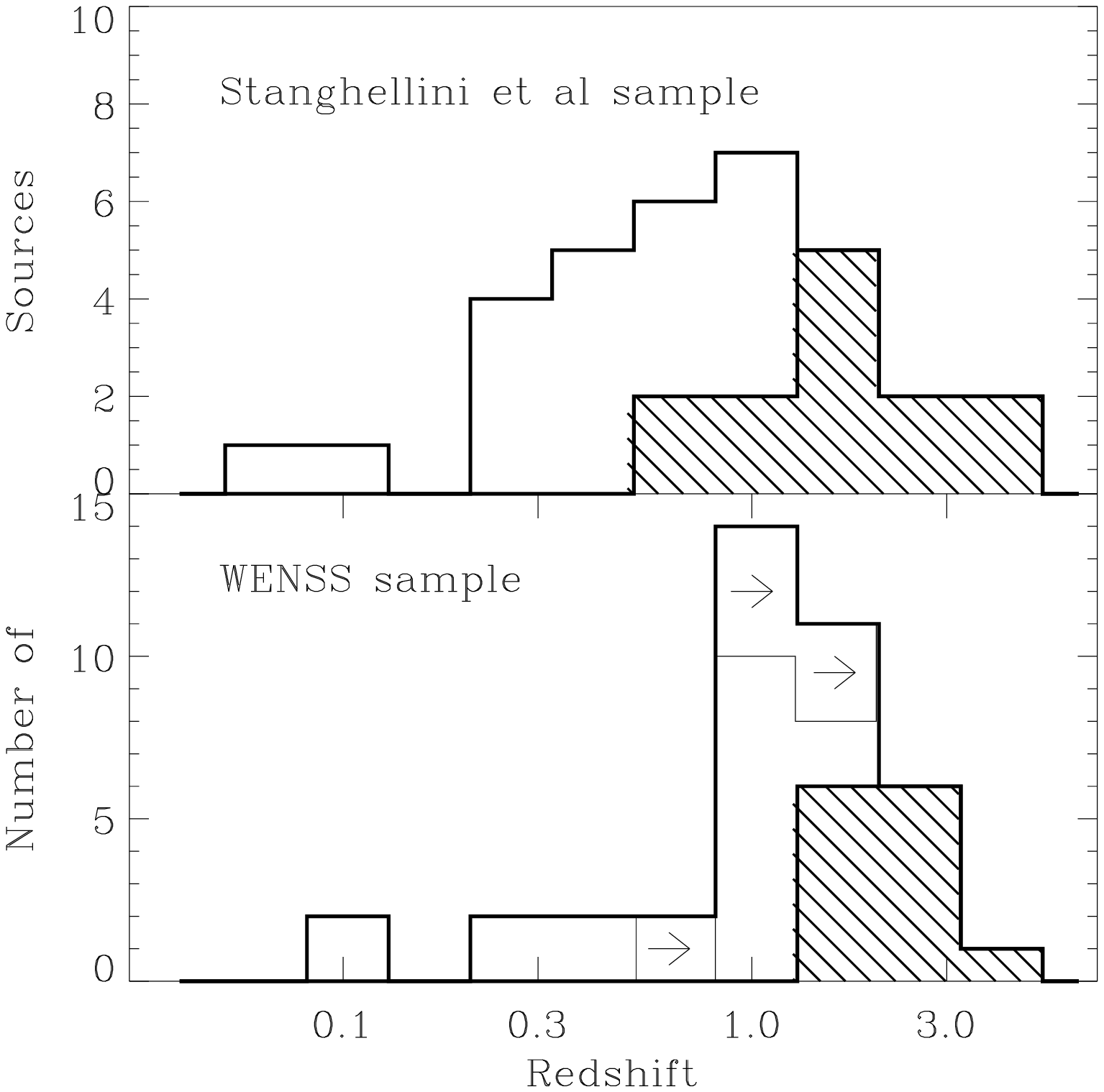,width=9cm}}
\caption{\label{zdist} The redshift distributions of the GPS galaxies (open)
and quasars (filled) in the radio-bright sample of Stanghellini et al. 
(upper panel) and the radio-faint sample (bottom panel). Note that there is 
a large uncertainty in the redshift distribution of the radio-faint galaxies.}
\end{figure}

\begin{figure}
\centerline{
\psfig{figure=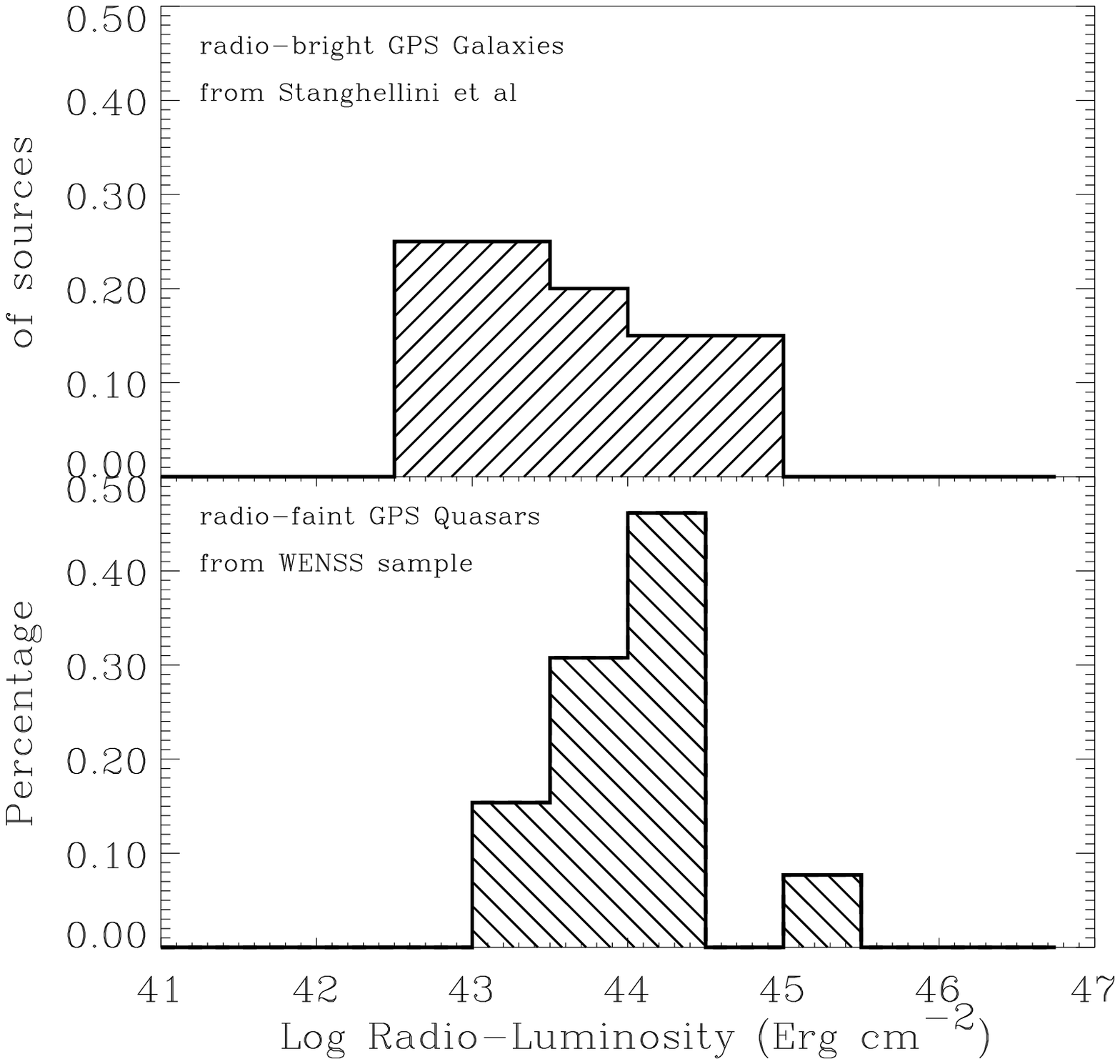,width=9cm}}
\caption{\label{lumdist} The radio luminosities of the GPS galaxies in the 
radio-bright sample are comparable to those of the GPS quasars in the 
radio-faint sample.}
\end{figure}

The optical counterparts of GPS sources at redshifts smaller than
unity are seldomly found to have broad emission lines (eg. 
Stanghellini et al. 1998). For some of those, the contrast between 
the stellar and non-stellar emission is low (eg. 1404+286; Osterbrock and Cohen, 1979 and B1622+6630;
this paper) which is why they are called Broad Line Radio Galaxies (BLRG) 
instead of quasars. Other `low' redshift GPS sources also have a substantial  
non-stellar contribution in their optical light as indicated by a power-law 
optical spectrum but have no broad emission lines (eg. 2342+821 
(Lawrence et al., 1996) and 0531+6121; this paper). 
However the large majority of low redshift
GPS sources ($\sim> 75\%$) are galaxies which show, apart from some
narrow emission lines, no evidence for a contribution of non-thermal
emission in the optical. 
High redshift GPS sources ($z>1.5$) are generally found to be 
genuine quasars (O'Dea 1990, O'Dea et al. 1991) with strong broad emission 
lines and no sign of the underlying host galaxies due to the high contrast
between  non-thermal nuclear emission and integrated star light. 

It is important to establish whether this dependence of optical host with
redshift is genuine, or whether it is due to selection 
effects; i) The known redshifts of GPS galaxies are likely to be biased towards
low redshift because it is increasingly difficult to measure their emission 
line features with redshift (at least up to $z\sim2$ at which Ly$\alpha$ and
CIV are shifted into the optical band). Note that the small fraction of low 
redshift 
BLRGs would be identified as galaxies if located at high redshift, since
their $H_{\alpha}/H_{\beta}$ ratio and stellar absorption lines indicate that
the non-stellar component is strongly absorbed and not visible at rest-frame 
UV wavelengths. Therefore, the classification of BLRG as galaxies and not as 
quasars does not bias the redshift distributions.
ii) The high redshift GPS quasars are
found to have substantially higher rest-frame radio peak frequencies than the 
GPS galaxies (Stanghellini et al. 1998, Snellen et al. 1998b). 
If a population of low redshift GPS quasars with similar peak frequencies as
the high redshift quasars exists, it would not have been included in the 
radio-bright samples due to their too high {\it observed} radio-peak 
frequencies. 

Both selection effects can be investigated with our faint sample. 
Due to the well established Hubble diagram for GPS galaxies (Snellen et al.
1996a), the redshifts of the GPS galaxies can be estimated using their
optical $R$ band magnitude. Although this method results in an increasing 
uncertainty with redshift, we believe it is sufficient to establish 
differences between the redshift distributions.
The faint GPS sample has been constructed in such a way that if an object
falls within the sample at $z=z_{gps}$, it also would have been in 
the sample if it was located at $z<z_{gps}$. This is due to the fact that
the sources are selected on the optically thick (inverted) part of their 
radio spectrum, and no high-frequency cut-off has been made.
 The shift in the peak
frequency to higher values due to a lower redshift would be compensated for 
by an increase in the peak flux density. 
For example, the GPS quasar B1945+6024 at z=2.70 is in our sample,
although it has an (undetermined) peak frequency larger than 15 GHz.
No low redshift GPS quasars are found in our sample, which makes it unlikely
that a significant population of low redshift GPS quasars is missing in 
the current GPS samples.

Knowing it to be free from major selection effects, the redshift distributions
 of the 
quasars and galaxies in the complete radio-bright GPS sample of 
Stanghellini et al (1998) and the faint sample can be 
compared. 
The radio luminosities of the objects in both samples
are calculated as the peak frequency times the peak flux density, and
assuming $H_0 = 50$ km/sec/Mpc and $q_0=0.5$. For only five (25\%) 
of the GPS galaxies in the Stanghellini et al. sample the redshifts were to be
estimated from their $R$ band magnitudes. This method had to be used
for the majority of the galaxies in the faint sample (80\%), and for 8
of these (26\%) only a lower limit of the redshift could be obtained.
All the redshifts of the quasars in the bright sample and 
70\% of those in the faint sample have been measured.
The radio luminosities of the GPS sources versus redshift are 
shown in figure \ref{zradio}. The diamonds indicate sources from the bright
sample and the squares indicate sources from the faint sample. Filled 
symbols are quasars and open symbols are galaxies.
Clearly in both samples the galaxies are found at low redshift and 
the quasars at high redshift. The GPS galaxies in the faint sample are
biased towards higher redshift than the galaxies in the radio-bright sample
(figure \ref{zdist}). In the faint sample, only 6 out of 27 galaxies (22\%) are
found at z$\leq$0.5, while in the radio-bright sample
12 out of 19 galaxies (63\%) are found at z $\leq$ 0.5. In addition, 
the faintest galaxies in the radio-faint sample are fainter than the 
faintest galaxies in the radio-bright sample. However, at present, it
is not clear whether this is due to
lower optical luminosities or higher redshifts of the host galaxies.
A small fraction of the faint and red optical identifications may turn
out to be quasars instead of galaxies. However, if quasars, they are most 
likely to be at even higher redshifts than the optically brighter quasars,
since the colours of quasars at $z>2$ become rapidly redder with redshift 
due to the absorption by intervening Ly$\alpha$ (eg. Hook et al., 1995).
Therefore, this would unlikely affect the observed difference in redshift
distribution.

The fact that in both samples the galaxies are found at low redshift and 
the quasars at high redshift, indicates 
that these different redshift-distributions are not caused by a 
radio-power effect; the higher the radio power, the larger the chance
to see the quasar nucleus.
Such an effect, that may be caused by a 
luminosity-dependent quasar opening-angle, could in a flux density limited
sample lead to the dependence of optical host with redshift. However, finding
a similar effect in both the faint and bright sample rules out this 
possibility. In figure \ref{lumdist} it is shown that the radio-luminosities 
of the GPS galaxies in the radio-bright sample are comparable to 
the luminosities of the GPS quasars in the radio-faint sample. 
This, combined with the fact that the quasar fraction is the same in both 
samples within the uncertainties, indicates that the galaxy-quasar 
difference is not a radio-power effect but a redshift effect.

The lifetime of a radio source is much smaller than cosmological time scales.
Therefore, if GPS quasars and galaxies are unified by orientation, their
different redshift distributions imply that the quasar opening angle is
a strong function of redshift. However, such a strong dependence of
opening angle with redshift needed to explain the data is not found for other
types of radio-loud AGN. 
Snellen et al. (1998c) show that if GPS 
galaxies were oriented towards us, that their overall radio spectrum would 
change from gigahertz peaked to flat spectrum and variable, due to Doppler 
boosting of the inner radio components. This is a strong indication that 
GPS galaxies and quasars are not unified by orientation and that they
form two distinct classes of objects which just happen to have the 
same observed radio-spectral properties. 

In this case, the GPS quasars may be a sub-class of flat spectrum 
sources in general, having physical parameters
at one end of the range seen in flat spectrum quasars. A 
model should account for their low flux density variability, peaked 
radio spectra, low extinction optical spectra, and low dispersion in the 
CIV-continuum relation as presented in section \ref{sectbaldwin}. For example,
the GPS quasars may have slower jets (due to a denser medium, O'Dea et al. 
1991) than 
the flat spectrum quasars in general, or their jets may be 
pointed more towards us resulting in a larger optical depth and possible
synchrotron self absorption. A detailed comparison of the redshift 
distributions and the radio-spectral properties of GPS and flat spectrum 
quasars, including variability studies, are needed to further investigate 
this hypothesis.

\section{Conclusions}

We have presented spectroscopic observations of a sample of faint 
Gigahertz Peaked Spectrum (GPS) radio sources drawn from the Westerbork 
Northern Sky Survey (WENSS). 
Redshifts have been determined for 19 (40\%) of the objects.
The optical spectra of the low redshift GPS galaxies show deep stellar 
absorption features, which confirms that their optical light is not 
significantly contaminated by AGN-related emission, but is dominated by a 
population of old ($>$9 Gyr) and metal-rich ($>$0.2 $[$Fe/H$]$) stars. 
The optical spectra of GPS sources identified with quasars are 
indistinguishable from those of flat spectrum quasars. Their blue colours
and strong 3000 \AA \ bump indicate low dust extinction, which is 
comparable to the spectra of core-dominated quasars, but clearly different 
from the spectra of Compact Steep Spectrum (CSS) quasars. 
The low dispersion in the Baldwin effect for GPS quasars indicates that
their optical continuum can only be mildly Doppler boosted.
The redshifts of the GPS quasars in our radio-faint sample are comparable
to those in the bright samples presented in the literature, and are
preferentially found at $z\sim 2-3$. The construction of the radio-faint sample
was such, that it is unlikely that a significant population of low redshift 
GPS quasars has been missed due to selection effects. We therefore claim that 
there is a genuine difference between the redshift distributions of 
GPS galaxies and quasars which, because it is present in both the 
radio-faint and bright samples, can not be due to a redshift-luminosity
degeneracy. It is unlikely that the GPS quasars and galaxies are unified
by orientation, unless the quasar opening angle is a strong function of 
redshift. We suggest that the GPS quasars and galaxies are not related and
just happen to have identical observed radio-spectral properties, 
and hypothesise that GPS quasars are a sub-class of flat spectrum quasars in 
general.

\section*{acknowledgements}

We thank the {\it Comite Cientifico International} (CCI) of the IAC
for the allocation of observing time. 
The Isaac Newton Telescope, and the William Herschel Telescope are
operated on the island of La Palma by the Isaac Newton Group in the Spanish
Observatorio del Roque de los Muchachos of the Instituto de Astrofisica de 
Canarias. We thank Richard McMahon and Neal Jackson for taking some of the 
spectra. This work was in part funded
through an NWO programme subsidy and by the European Commission under
contracts SCI*-CT91-0718 (The Most Distant Galaxies) and 
ERBFMRX-CT96-086 (Formation and Evolution of Galaxies), and
ERBFMRX-CT96-0034 (CERES).

{}


\begin{thebibliography}{}
\bibitem{} Baker J.C. and Hunstead R.W., 1995, {\it Astrophys. J.}, L95
\bibitem{} Baldwin J.A., 1977, {\it Astrophys. J.}, {\bf 214}, 769 
\bibitem{} Baldwin J.A., Wampler E.J. and Gaskell C.M., 1989,
           {\it Astrophys. J.}, {\bf 338}, 630 
\bibitem{} Dressler A., Lynden-bell D., Burstein D., Davies R.L., Faber S.M., 
           Terlevich, R., Wegner G., 1987, {\it Astrophys. J.}, {\bf 313}, 42 
\bibitem{} Fanti R., Fanti C., Schilizzi R.T., Spencer R.E., Nan Rendong, 
             Parma P., Van Breugel W.J.M., Venturi T., 1990, 
             {\it Astron. Astrophys.}, {\bf 231}, 333
\bibitem{} Fanti C., Fanti R., Dallacasa D., Schilizzi R.T., Spencer R.E.,
             Stanghellini C., 1995, {\it Astr. \& Astrophys.}, {\bf 302}, 317
\bibitem{} Hook I.M., McMahon R.G., Patnaik A.R., Browne I.W.A., 
           Wilkinson P.N., Irwin M.J., Hazard C., 1995, 
           {\it Mon. Not. R. Astr. Soc.}, {\bf 273}, L63
\bibitem{} Mushotzky R. and Ferland G.J., 1984, {\it Astrophys. J.}, {\bf 278}, 558 
\bibitem{} Kinney A.L., Huggins P.J., Bregman J.N., Glassgold A.E., 1985, 
           {\it Astrophys. J.}, {\bf 291}, 135 
\bibitem{} Lawrence C.R., Zucker J.R., Readhead A.C.S., Unwin S.C., 
           Pearson T.J., Xu W.,, 1996, {\it Astrophys. J. Suppl.}, {\bf 107}, 541
\bibitem{} O'Dea C.P., 1990, {\it Mon. Not. R. Astr. Soc.}, {\bf 245}, 20
\bibitem{} O'Dea C.P., Baum S.A., Stanghellini C., 1991, {\it Astrophys. J.},
             {\bf 380}, 66
\bibitem{} O'Dea C.P., Stanghellini C., Baum S.A., Charlot S., 1996,
           {\it Astrophys. J.}, {\bf 470}, 806
\bibitem{} O'Dea C.P., Baum S.A., 1997, {\it Astron. J.}, 
          {\bf 113}, 148
\bibitem{} O'Dea C.P., 1998, P.A.S.P., {\bf 110}, 493
\bibitem{} Osmer P.S., Porter A.C., Green R.F., 1994,
           {\it Astrophys. J.}, {\bf 436}, 678
\bibitem{} Osterbrock and Cohen, 1979, {\it Mon. Not. R. Astr. Soc.}, {\bf 187}, 61
\bibitem{} Owsianik I. and Conway J., 1998, {\it Astr. \& Astrophys.}, 
{\bf 337}, 69 
\bibitem{} Readhead A.C.S., Taylor G.B., Xu W., Pearson T.J., Wilkinson P.N.,
             Polatidis A.G., 1996, {\it Astrophys. J.}, {\bf 460}, 612
\bibitem{} Rengelink R.B., Tang Y., de Bruyn A.G., Miley G.K., Bremer M.N., 
             R\"ottgering H.J.A., Bremer M.A.R., 1997, 
             {\it Astron.  Astrophys. Suppl.}, {\bf 124}, 259 
\bibitem{} Snellen I.A.G., Bremer M.N., Schilizzi R.T., Miley G.K., 
             van Ojik R., 1996a, {\it Mon. Not. R. Astr. Soc.}, {\bf 279}, 1294
\bibitem{} Snellen I.A.G., Bremer M.N., Schilizzi R.T., Miley G.K., 1996b,
             {\it Mon. Not. R. Astr. Soc.}, {\bf 283}, 123
\bibitem{} Snellen I.A.G., {\it PhD thesis} 1997, Leiden Observatory
\bibitem{} Snellen I.A.G., Schilizzi R.T., de Bruyn A.G., Miley G.K.,
           Rengelink R.B., R\"ottgering H.J.A., Bremer, M.N., 1998a,
           {\it Astr. \& Astrophys. Suppl.}, {\bf 131}, 435 
\bibitem{} Snellen I.A.G., Schilizzi R.T., Bremer M.N., de Bruyn A.G., Miley G.K., R\"ottgering H.J.A.,
           McMahon R.G., P\'erez Fournon I., 1998b, {\it Mon. Not. R. Astr. Soc.}, {\bf 301}, 985
\bibitem{} Snellen I.A.G., Schilizzi R.T., de Bruyn A.G., Miley G.K., 1998c,
          {\it Astr. \& Astrophys.}, 333, 70
\bibitem{} Stanghellini C., O'Dea C.P., Baum S.A., Dallacasa D., Fanti R., Fanti C., 1997, 
           {\it Astron. \& Astrophys.}, {\bf 325}, 943
\bibitem{} Stanghellini C., O'Dea, C. P., Dallacasa, D., Baum, S.A., Fanti, R.,Fanti, C., 1998, {\it Astron. \& Astrophys. Suppl.}, {\bf 131}, 303 
\bibitem{} De Vries W.H., Barthel P.D., O'Dea C.P., 1997,
             {\sl Astronomy and Astrophysics}, {\bf 321}, 105
\bibitem{} Wilkes B.J., Wright A.E., Jauncey D.L., Peterson B.A., 1983,
           {\it Proceedings Astr. Soc. of Australia}, {\bf Vol. 5}, no. 1, p2
\bibitem{} Wills B.J., Netzer H., Brotherton M.S., Han Mingsheng, Wills D., 
           Baldwin J.A., Ferland G.J., Browne I.W.A., 1993, {\it Astrophys. J},
           {\bf 410}, 534
\bibitem{} Worthey G., 1994, {\it Astrophys. J. Suppl.}, {\bf 95}, 107
\vfill
\end{thebibliography}
\end{document}